\documentclass[journal]{IEEEtran}
\usepackage{amsbsy}
\usepackage{bm}
\usepackage{color}
\usepackage{booktabs}
\usepackage{algorithm}
\usepackage{algorithmic}
\usepackage{amssymb}
\usepackage{amsmath}
\usepackage{amsthm}
\usepackage{comment}
\usepackage{hyperref}
\usepackage{tikz}
\usetikzlibrary{graphs, positioning, quotes, shapes.geometric}

\newtheorem{thm}{Theorem}

\newtheorem{prop}[thm]{Proposition}

\newcommand{\br}{{\mathbb R}}

\newcommand{\diag}{{\rm diag}}

\newcommand{\prf}{\par{\bf Proof. }}
\newcommand{\bbox}{\rule{2mm}{3mm}}
\newcommand{\rank}{{\rm rank}}
\newcommand{\inv}{{\rm INV}}
\newcommand{\lp}{{\rm L}}
\newcommand{\hp}{{\rm H}}
\newcommand{\s}{{\rm S}}

\newcommand{\bfu}{\mathbf{u}}
\newcommand{\bfv}{\mathbf{v}}

\newcommand{\bfh}{\mathbf{h}}
\newcommand{\bfx}{\mathbf{x}}
\newcommand{\bfy}{\mathbf{y}}
\newcommand{\bfw}{\mathbf{w}}
\newcommand{\bfA}{\mathbf{A}}
\newcommand{\bfC}{\mathbf{C}}
\newcommand{\bfD}{\mathbf{D}}

\newcommand{\bfH}{\mathbf{H}}
\newcommand{\bfI}{\mathbf{I}}
\newcommand{\bfK}{\mathbf{K}}

\newcommand{\bfL}{\mathbf{L}}
\newcommand{\bfG}{\mathbf{G}}
\newcommand{\bfU}{\mathbf{U}}

\newcommand{\mcA}{\mathcal{A}}
\newcommand{\mcB}{\mathcal{B}}

\title{Spline-Like Wavelet Filterbanks with Perfect Reconstruction on Arbitrary Graphs}

\begin{document}

		This article has been accepted for publication in IEEE Transactions on Signal and Information Processing over Networks. This is the author's version which has not been fully edited and content may change prior to final publication. Citation information: DOI 10.1109/TSIPN.2023.3264993.
	\newpage
%\normalem 	
	\author{Junxia You and Lihua Yang
		
		\thanks{This is supported by National Natural Science Foundation of China 
			(No. 12171488) and Guangdong Province Key Laboratory of Computational Science 
			at the Sun Yat-sen University (2020B1212060032). Note: Codes for this paper are available at https://github.com/ConquerBroccoli/matlabcodes/tree/codes-for-thesis. Please contact Junxia You for further questions about the codes.}
		\thanks{Junxia You is with School of Mathematics, Sun Yat-sen University, Guangzhou, China (e-mail: youjx3@mail2.sysu.edu.cn).} 
		\thanks{Lihua Yang is with School of Mathematics, Sun Yat-sen University, Guangzhou, China and Guangdong Province Key Laboratory of Computational Science (e-mail: mcsylh@mail.sysu.edu.cn). He is the corresponding author.}
	}
	
	\markboth{IEEE TRANSACTIONS ON SIGNAL AND INFORMATION PROCESSING OVER NETWORKS, 2022}%
	{Junxia You, Lihua Yang: Spline-Like Wavelet Filterbanks with Perfect Reconstruction on Arbitrary Graphs}
	
	\maketitle	
	\begin{abstract}
		In this work, we propose a class of spline-like wavelet filterbanks for graph signals. These filterbanks possess the properties of critical sampling and perfect reconstruction. The analysis filters are localized in the graph domain because they are polynomials in the normalized adjacency matrix of the graph. We generalize the spline-like filters in the literature so that the lowpass filter and the highpass filter can respectively remove the $s$ highest frequency components and the $r$ lowest frequency components of the signal, where $r$ and $s$ are hyperparameters specified by the users. Optimization models are formulated for the analysis filters to approximate the desired responses. Experimental results demonstrate the good locality and denoising ability of the proposed filterbanks.
	\end{abstract}

%--------------------引言------------------
\section{Introduction}

In recent years, complex data analysis is widely concerned. In many applications, data structures such as social networks, sensor networks and biological networks can be modelled as graphs, and the data residing on these graphs are called graph signals.
With the rise of big data science, theoretical and applied research on graph signal processing (GSP) becomes increasingly important. Researchers are working to extend the theory and methods in classical signal processing to GSP. Theories about graph Fourier transform, graph filters, graph wavelets and Multiresolution analysis (MRA) on graph signals are developed \cite{Sandryhaila-2013b,yang2021graph,gavish2010multiscale,Shuman2016Amultiscale}. In terms of application, GSP methods are also widely used in such as point clouds analysis \cite{hu2022exploring,liu2022point}, deep neural networks and computer vision \cite{liu2021spatiotemporal}.
However, due to the irregularity of graph structure, there are still many challenges in this field.

Wavelet analysis of graph signals is an important topic in GSP.	
Researchers have developed different types of graph wavelets. In \cite{crovella2003graph}, Crovella and Kolaczyk constructed a series of compactly supported simple functions on each neighbourhood of every vertex as graph wavelet functions. 
Coifman and Maggioni proposed the concept of diffusion wavelets in \cite{coifman2006diffusion}.
	Gavish et al. \cite{gavish2010multiscale} first constructed multiscale wavelet-like orthonormal bases on hierarchical trees. Hammond et al. \cite{hammond2011wavelets} constructed wavelet transforms in the graph domain based on the spectral graph theory. In follow-up work, they designed an almost tight wavelet frame based on the polynomial filters \cite{tay2017almost}. In \cite{Shuman2016Amultiscale}, Shuman et al. proposed a modular framework--a multiscale pyramid transform for graph signals.
	All these wavelets are not critically sampled, the output has more components than the input signal, which leads to the waste of space for storing redundant information. 
Critically sampled wavelet filterbanks have also been proposed in many works. Narang and Ortega developed the two channel filterbanks composed of graph quadrature mirror filters and the compact support biorthogonal filterbanks in \cite{narang2012perfect,narang2013compact}. Ekambaram et al. proposed the spline-like filterbanks in \cite{Ekambaram2015Spline}. The exponential spline filterbanks on circulant graphs are proposed by Kotzagiannidis and Dragotti in \cite{2016Splines}, and the modified spline filterbanks are proposed by Miraki et al. in \cite{miraki2021modified} and \cite{miraki2021spline}. Especially, the shceme proposed in \cite{miraki2021spline} utilizes the spectral domain sampling method proposed in \cite{tanaka2018spectral}.

	The classical wavelets can capture local information of signals in the time domain, i.e., each sample of the transformed signal is computed by using the samples from a small neighbourhood of the original signal. This property enables wavelets to capture the details of the signal. Thus, we are interested in the spline-like filterbanks proposed in \cite{Ekambaram2015Spline}, since the analysis filters are polynomials in 
	the normalized adjacency matrix of the graph, which leads to the locality of filters in the graph domain. 
	
	The authors of \cite{Ekambaram2015Spline} provide results on the perfect reconstruction property of their proposed spline-like filterbanks, and formulate optimization models to obtain the desired filter responses. The filterbanks have the advantage of critical sampling, and the analysis filters are well localized in the graph domain.
	However, the lowpass filter they designed cannot remove the highest frequency component of the signal unless it is a degree-$1$ polynomial in the normalized adjacency matrix and the graph is bipartite, 
	as discussed later in Section \ref{sec:related}. This will impair the denoising ability of the filterbanks. 
	Therefore, we extend their work to enable the filterbanks with better denoising capability. The novelty and main contributions of this paper are summarized as follows.

	We propose a class of spline-like filterbanks in which the lowpass (highpass) filter can remove more than one high-frequency (low-frequency) components of the signals. A perfect reconstruction theorem is established where the sampling pattern is required to meet some mild conditions and an algorithm is proposed to obtain the effective sampling pattern.  Similarly, optimization problems are formulated for the analysis filters to approximate the desired frequency responses. 
	We also construct filterbanks based on the non-normailzed adjacency matrices, which is useful in some applications that require the highpass filter to eliminate the direct current (DC) signal. Besides,  through a counterexample we point out a small flaw in the perfect reconstruction theorem in \cite{Ekambaram2015Spline} and give a correction.

	This paper is organized as follows: in Section \ref{sec:background}, we introduce some basic concepts related to the graph filterbanks and introduce the design in \cite{Ekambaram2015Spline} to motivate our work. In Section \ref{sec:3}, we describe the proposed generalized spline-like filterbanks, and provide sufficient conditions for the filterbanks to be perfectly reconstructed. Besides, we give an algorithm to obtain sampling patterns that satisfy the perfect reconstruction conditions and formulate optimization models for the filters to approximate desired responses. In Section \ref{sec:expe}, experiments are conducted to demonstrate the effectiveness of the proposed filterbanks compared to the related work. Finally, we make a conclusion and discuss the limitation and future work in Section \ref{sec:conclusion}.

	%----------------------------正文-----------------------------
	\section{Preliminary}\label{sec:background}
	
	\subsection{Notations}
	We use bold letters for matrices and vectors, calligraphic capital letters for sets, and normal letters for scalars.
	
	The $i$-th entry of a vector $\bfx$ is denoted by $x_i$ or $\bfx(i)$. 
	The $(i,j)$-th entry of a matrix $\bfA$ is denoted by ${\bfA}(i,j)$. Assume that $\mathcal{I}_1,\mathcal{I}_2$ are two subsets of $\{1,...,N\}$, then $\bfA(\mathcal{I}_1,\mathcal{I}_2)$ denotes the submatrix consisting of entries of $\bfA$ whose row indices are in $\mathcal{I}_1$ and column indices are in $\mathcal{I}_2$.
	Let $\bfI_N$ represent the identity matrix of order $N$ and $\mathbf{1}, \mathbf{0}$ respectively represent the all-ones vector and the null vector whose sizes can be seen from the context. 
	
	The superscript $^\top$ denotes transposition. $\diag(\cdot)$ maps a vector to a diagonal matrix, or a matrix to its main diagonal vector. The infinity norm and $2$-norm of a vector $\bfx\in\br^N$ are defined as $\|\bfx\|_\infty=\max_{1\leq i\leq N}|x_{i}|$ and $\|\bfx\|_2=(\sum_{i=1}^{N}|x_i|^2)^{\frac{1}{2}}$, respectively. $\bfx>(\geq)0$ means that all entries of $\bfx$ are positive (non-negative). The $2$-norm of a matrix, denoted by $\|\bfA\|_2$, is defined as the largest singular value of $\bfA$. The cardinality of a set $\mathcal{V}$ is written as $|\mathcal{V}|$. 
	
	\subsection{Graph and Graph Fourier Transform}
	\par 
	A graph can be denoted as $\mathcal{G}=(\mathcal{V},\mathcal{E},\bfA)$ with vertex  set $\mathcal{V}=\{1,...,N\}$, edge set $\mathcal{E}=\{(i,j)|~i\sim j\}$ and adjacency matrix $\bfA$, where $i\sim j$ means that vertices $i$ and $j$ are connected.
	We only consider connected, undirected and weighted graphs without self-loops or multiple edges in this paper. 
	The elements of $\bfA$ indicate the adjacency relationship of pairs of vertices such that $\bfA(i,j)>0$ if $(i,j)\in\mathcal{E}$ and $\bfA(i,j)=0$ otherwise.
	Let $\bfD=\diag(d_1,...,d_N)$ denote the degree matrix of $\bfA$, where $d_i=\sum_{j=1}^{N}\bfA(i,j)$ is the degree of vertex $i$. 
	
	Due to the connectivity of $\mathcal{G}$, $\bfD$ is non-singular. Thus, we can define the symmetric normalized adjacency matrix as $\bfA^\s=\bfD^{-\frac{1}{2}}\bfA\bfD^{-\frac{1}{2}}$. 
	Correspondingly, the symmetric normalized Laplacian matrix of $\mathcal{G}$
	is defined as $\bfL^\s=\bfI_N-\bfA^\s$ \cite{chung1997spectral}. 
	Since ${\bfL^\s}$ is real symmetric and positive semi-definite,
	there exists a set of orthonormal eigenvectors $\{\bfu_l\}^N_{l=1}$ and real eigenvalues $0=\lambda_1<\lambda_2\le\cdots\le\lambda_N$
	such that $\bfL^\s={\bfU}\boldsymbol{\Lambda}{\bfU}^\mathrm{\top}$, where $\bfU=(\bfu_1,...,\bfu_N)$ and $\boldsymbol{\Lambda}=\diag(\lambda_1,...,\lambda_N)$.
	Obviously, the eigendecomposition of $\bfA^\s$ can be written as $\bfA^\s=\bfU\diag(\xi_1,...,\xi_N)\bfU^\top$ with $\xi_i=1-\lambda_i$. In the rest of the paper, $\{\bfu_i\}_{i=1}^N$ and $\bfU$ will always denote the eigenvectors and the corresponding eigenmatrix of $\bfL^\s$, and $\bfu_i$ is called the $i$-th \textbf{Fourier basis vector} of frequency $\lambda_i$ which increases as $i$ goes from $1$ to $N$.

	A graph signal $x:\mathcal{V}\to\br$ is a function defined on the vertices of the graph. If the labels of the vertices are fixed, the signal can also be written as a vector 
	$\bfx\in\br^N$. In this paper, we define the graph Fourier transform (GFT) of signal $\bfx$ as $\hat{{\bfx}}={\bfU}^{\top}{\bfx}$ \cite{shuman2013emerging}.
	Thus, $\bfx$ can be represented as $\bfx=\sum_{l=1}^{N}\hat{\bfx}(l)\bfu_l$, and $\hat{\bfx}(l)\bfu_l$ is referred to as the component of $\bfx$ with frequency $\lambda_l$.

	\subsection{Two-Channel Filterbanks and Related Work}\label{sec:related}
\begin{figure}[hbtp]
		\begin{center}
			\tikzstyle{basic} = [rectangle, minimum width=0.8cm,minimum height=0.5cm,text centered,draw=black]
			\tikzstyle{du} = [rectangle, rounded corners, minimum width=0.8cm,minimum height=0.5cm,text centered,draw=black]
			\tikzstyle{arrow} = [thick,->,>=stealth]
			\begin{tikzpicture}[node distance=10pt]
				\node[rounded corners,draw=black] (x)  {$\bfx$};
				%Lowpass
				\node[basic,above right=15pt of x,yshift=-0.4cm]   (DL)  {$\bfH_\lp$};
				\node[du,right=12pt of DL]                (SL)  {$\downarrow_\lp$};
				\node[draw, right=12pt of SL]     					(pro1)  {\small{process}};
				\node[du,right=12pt of pro1]						(SLU){$\uparrow_\lp$};
				%Highpass
				\node[basic,below right=15pt of x,yshift=0.4cm]            (DH)  {$\bfH_\hp$};
				\node[du,right=12pt of DH]                        (SH)  {$\downarrow_\hp$};
				\node[draw, right=12pt of SH]     					(pro2)  {\small{process}};
				\node[du,right=12pt of pro2]						(SHU){$\uparrow_\hp$};
				\node[basic,draw=black,right=155pt of x]						(R){$\bfH_{_{\inv}}$};			
				\node[rounded corners, draw=black,right=10pt of R]  (y)     {$\bfy$};
				
				%---------
				\draw [arrow] (x) |- (DL);
				\draw [arrow] (x) |- (DH);
				\draw [arrow] (DL) -- (SL);
				\draw [arrow] (DH) -- (SH);
				\draw [arrow] (SL) -- (pro1);
				\draw [arrow] (pro1) -- (SLU);
				\draw [arrow] (SH) -- (pro2);
				\draw [arrow] (pro2) -- (SHU);
				\draw [arrow] (SLU) -| (R);
				\draw [arrow] (SHU) -| (R);
				\draw [arrow] (R) -- (y);
			\end{tikzpicture}
		\end{center}
		\caption{A two-channel filterbank.}\label{FB}
		\label{fig:2-channel-banks}
	\end{figure}
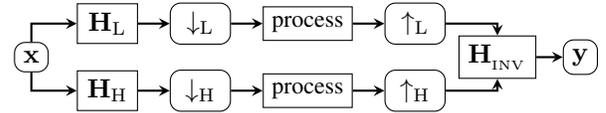

	A two-channel filterbank is shown in Figure  \ref{fig:2-channel-banks}. It is a collection of filters and samplers. The filters $\bfH_\lp,\bfH_\hp$ are called analysis filters and the filter $\bfH_{\inv}$ is called synthesis filter, where the subscript $_\lp$ represents lowpass (LP) and $_\hp$ represents highpass (HP). The  downsampler and the upsampler are denoted by $\downarrow$ and $\uparrow$ respectively.

	%----------------------------------------------------

	Given a graph signal $\bfx\in\br^N$, the analysis filters $\bfH_\lp,\bfH_\hp$ attenuate the high and low frequency components of $\bfx$ respectively. After that, the filtered signal in each channel will be downsampled to produce signals $\bfy_\lp=(\downarrow_\lp)\bfH_\lp \bfx$ and $\bfy_\hp=(\downarrow_\hp)\bfH_\hp \bfx$. 
	If the sum of lengths of $\bfy_\lp$ and $\bfy_\hp$ equals $N$, the filterbank is said to be critically sampled. In this case, we can define a sampling matrix $\bfK=\diag(k_1,...,k_N)$ with $k_i\in\{1,-1\},\forall i=1,..,N$ such that $\bfy_\lp$ is a subvector of $\bfH_\lp\bfx$ with indices in $\{i|k_i=1\}$ and $\bfy_\hp$ is a subvector of $\bfH_\hp\bfx$  with indices in $\{i|k_i=-1\}$.
	
	After downsampling, the signals may be encoded for transmission or storage, which may result in loss of information. To construct a perfect reconstruction filterbank such that $\bfy=\bfx$, we omit the processing stage, i.e., upsample the signals immediately after downsampling. Thus, we have
	\begin{equation}\label{eq:TCFB}
		\bfy=\bfH_{\inv}\big[\frac{1}{2}(\bfI_N+\bfK)\bfH_\lp+\frac{1}{2}(\bfI_N-\bfK)\bfH_\hp\big]\bfx.
	\end{equation}
	The filterbank is perfectly reconstructed if and only if (iff)
	\begin{equation}\label{eq:pr}
		\bfH_{\inv}\big[\frac{1}{2}(\bfI_N+\bfK)\bfH_\lp+\frac{1}{2}(\bfI_N-\bfK)\bfH_\hp\big]=\bfI_N.
	\end{equation}

	\par 
	Inspired by the classical  first-order spline filters, the authors of \cite{Ekambaram2015Spline} designed a class of spline-like analysis filters for the two-channel filterbanks on graphs, which are
	\begin{align}\label{spline:general}
		\begin{split}
			&\bfH_\lp^\s=\frac{1}{2}\Big(\bfI_N+\sum_{l=1}^Jw_l(\bfA^\s)^l\Big),\\
			&\bfH_\hp^\s=\frac{1}{2}\Big(\bfI_N-\sum_{l=1}^Jw_l(\bfA^\s)^l\Big),
		\end{split}
	\end{align}
	where the weights $w_1,...,w_J$ are positive scalars. The corresponding filter responses are given as
	\begin{align}
		\begin{split}
			&\bfh_\lp^\s(i)=\frac{1}{2}\Big(1+\sum_{l=1}^Jw_l\xi_i^l\Big),\\
			&\bfh_\hp^\s(i)=\frac{1}{2}\Big(1-\sum_{l=1}^Jw_l\xi_i^l\Big),
		\end{split}~~i=1,...,N.
	\end{align}
	The weights give us the flexibility to optimize the filter responses to the desired responses. Degree $J$ is a hyperparameter to be specified. The smaller $J$ is, the better the locality of filters in the graph domain (vertex domain).
	Let us take an example to illustrate the locality of filters in the graph domain. When $J=1$, we have
	\begin{align}
		(\bfH_\lp^\s\bfx)(i)=\frac{1}{2}\Big( x_i+w_1\sum_{\bfA^\s(i,j)>0}x_j\bfA^\s(i,j)\Big).
	\end{align}
	It is clear that $(\bfH_\lp^\s\bfx)(i)$ is determined by the entries of $\bfx$ located on the one-hop neighbourhood of vertex $i$. A $k$-hop neighbourhood of vertex $i$ is defined as $\{j|~\big[\sum_{l=1}^{k}(\bfA^\s)^l\big](i,j)>0\}$.

The authors of \cite{Ekambaram2015Spline} provided sufficient conditions for perfect reconstruction of the filterbanks with analysis filters defined in \eqref{spline:general}.
	%-------------------------------------------------------
	\begin{thm}\cite{Ekambaram2015Spline}\label{thm:cite}
		For any connected graph, the spline filters defined in \eqref{spline:general} form a critically-sampled, perfect reconstruction filterbank for any downsampling pattern, as long as the weights satisfy one of the following properties:
		\begin{align}\label{thm:cond:cite}
			\begin{cases}
				w_l>0,~~l=1,...,J,\\
				\sum_{l=1}^Jw_l=1,
			\end{cases}
			~\mbox{or}~\Big|\sum_{l=1}^{J}w_l\xi_i^l\Big|>1,
		\end{align}
		for any $i=1,...,N$, where $\{\xi_i\}_{i=1}^N$ are the eigenvalues of $\bfA^\s$.
	\end{thm}
	%-------------------------------------------------------
	\par
	We point out that the theorem is not mathematically accurate in the extreme case where $J=1$ and $\bfK=\bfI_N$, i.e., the downsampling pattern does not retain any highpass components.
	A counterexample is given below. When $J=1$, there holds $w_1=1$ and thus $\bfH_\lp^\s=\frac{1}{2}(\bfI_N+\bfA^\s)$.
		If $-1$ is an eigenvalue of $\bfA^\s$ (i.e., the graph is bipartite \cite{chung1996lectures}), then $0$ is an eigenvalue of $\bfH_\lp^\s$. In this case,
		\begin{align}
			\frac{1}{2}(\bfI_N+\bfK)\bfH_\lp^\s+\frac{1}{2}(\bfI_N-\bfK)\bfH_\hp^\s=\bfH_\lp^\s,
		\end{align}
		and $\bfH_\lp^\s$ is irreversible. Consequently, there is no matrix $\bfH_{\inv}^\s$ satisfying the perfect reconstruction equation \eqref{eq:pr}.
	However, the conclusion of the theorem can be proven correct if the downsampling pattern preserves at least one lowpass component and one highpass component, i.e., $\bfK\neq \pm \bfI_N$.

	By Theorem \ref{thm:cite}, one can formulate an optimization model to optimize the  weights $w_1,...,w_J$ to obtain the desired filter responses while satisfying the conditions for perfect reconstruction. For example, a least-square formulation is as follows \cite{Ekambaram2015Spline}:
	\begin{align}\label{opt:cite}
		\begin{split}
			\underset{\mathbf{w}\in\br^J}{\mbox{min}}~~~&\|\bfH^{\text{des}}-\bfH^\s_\lp\|_2\\
			\text{s.t.}~~~&\mathbf{w}^\top \mathbf{1}_J=1,\\
			\quad~~~&\bfw>0,
		\end{split}
	\end{align} 
	where $\bfw=[w_1,...,w_J]^\top$ and $\bfH^{\text{des}}$ is a desired lowpass filter.
	Figure 6 in \cite{Ekambaram2015Spline} shows an example of lowpass and highpass spline-like filter responses on the Tapir dataset, where $J=10$ and $\bfH^{\text{des}}$ is an ideal lowpass filter. We notice that the filter responses in the figure have a wide range from $0$ to $10^{15}$ and the lowpass response approaches $0$ near zero frequency, making it a bandpass filter instead of a lowpass filter. In addition, the presented ``highpass'' filter is not actually highpass since the filter response has high amplitude in the low frequency region and low amplitude in the high frequency region.
	
	We find that under the first set of conditions of Theorem \ref{thm:cite}, the highpass filter $\bfH_\hp^\s$ has a zero response to the lowest frequency Fourier basis vector $\bfu_1$
		(while the second set of conditions does not guarantee this), but the lowpass filter $\bfH_\lp^\s$ has a non-zero response to  $\bfu_N$, the highest frequency Fourier basis vector, unless $J=1$ and the graph is bipartite. This is because under the first set of conditions, the LP filter response $\bfh_\lp^\s(i)=0$ iff $\sum_{l=1}^{J}w_l\xi_i^l=-1$, which can only be achieved when $J=1$ and $\xi_i=-1$. However, there exists an eigenvalue $\xi_i=-1$ iff the graph is bipartite.
This fact weakens the denoising ability of the analysis filters 
for non-bipartite graphs. 
In Section \ref{sec:3}, we improve their design so that the LP filter has zero responses to the $s$ Fourier basis vectors $\{\bfu_i\}_{i=N-s+1}^N$ with the highest frequencies and the HP filter has zero responses to the $r$ Fourier basis vectors $\{\bfu_i\}_{i=1}^r$ with the lowest frequencies, where $r,s\ge1$ are hyperparameters specified by the users.

	%----------------------------------------------------------------
	\section{Generalization}\label{sec:3}
	\subsection{Main Theorem}\label{subsec:main}
	Hereafter, we consider analysis filters of the form:
	\begin{align}\label{filters}
		\begin{cases}
			&\bfH_\lp=\frac{1}{2}(\bfI_N+\sum_{l=1}^{J}w_l(\bfA^\s)^{l-1})\\
			&\bfH_\hp=\frac{1}{2}(\bfI_N-\sum_{l=1}^{J}w_l(\bfA^\s)^{l-1})
		\end{cases},
	\end{align}
	where $\bfw=[w_1,...,w_J]^\top\in\br^J$ and $J\geq2$. 
	Recall that $\bfA^\s=\bfU\diag(\xi_1,...,\xi_N)\bfU^\top$. For simplicity, denote 
	\begin{align*}
		&\bfG=\sum_{l=1}^{J}w_l(\bfA^\s)^{l-1},~\gamma_i=\sum_{l=1}^{J}w_l\xi_i^{l-1},\\
		&\boldsymbol{\Gamma}=\diag(\gamma_1,...,\gamma_N).
	\end{align*}
	Then $\bfG=\bfU\boldsymbol{\Gamma}\bfU^\top$. Similar to the scheme proposed in \cite{Ekambaram2015Spline},
	we will optimize the weights $\bfw$ for the desired filter responses while maintaining the perfect reconstruction property of the filterbank. 
	
There are some discussions before giving the formulation. Given a sampling matrix $\bfK$ and analysis filters $\bfH_\lp,\bfH_\hp$, the filterbank is perfectly reconstructed iff the synthesis filter $\bfH_{\inv}$ exists such that \eqref{eq:pr} holds. A simple calculation shows that
	\begin{align}
		\begin{split}
			\frac{1}{2}(\bfI_N+\bfK)\bfH_\lp+\frac{1}{2}(\bfI_N-\bfK)\bfH_\hp=\frac{1}{2}(\bfI_N+\bfK\bfG).
		\end{split}
	\end{align}
	Thus, $\bfH_{\inv}$ exists iff $\bfI_N+\bfK\bfG$ is invertible, in which case $\bfH_{\inv}=2(\bfI_N+\bfK\bfG)^{-1}$. 
	Next, we will design $\bfK$ and $\bfG$ so that the following two conditions are satisfied:
	\begin{itemize}
		\item [($1^\circ$)] $\bfI_N+\bfK\bfG$ is invertible;
		\item[($2^\circ$)] $\bfH_\lp\bfu_i=\mathbf{0},i=N-s+1,...,N$ and $\bfH_\hp\bfu_i=\mathbf{0}$, $i=1,...,r$, where $r\geq1,s\geq1$ are hyperparameters.
	\end{itemize}

\begin{figure}[hbtp]
		\begin{center}
			\tikzstyle{basic} = [rectangle, minimum width=3.6cm,minimum height=1.3cm,text centered,draw=black]
			\tikzstyle{arrow} = [thick,->,>=stealth]
			\begin{tikzpicture}[node distance=10pt]
				\node[basic,draw=black, align=center,font=\scriptsize] (s1)  {Specify the hyperparameters \\ $(r,s,J)$};
				\node[basic,right=15pt of s1, align=center,font=\scriptsize]   (s2)  {Solve the optimization \\ models to obtain $\bfw$};
				\node[basic,below=15pt of s2, align=center,font=\scriptsize]   (s3)  {Compute a partition $\{\mcA,\mcB\}$  \\ by Algorithm \ref{alg:A_B} and \\  obtain the corresponding $\bfK$};
				\node[basic,left=15pt of s3, align=center,font=\scriptsize]   (s4)  {Calculate the synthesis \\ filter as \\ $\bfH_{\inv}=2(\bfI_N+\bfK\bfG)^{-1}$};
				
				%---------
				\draw [arrow] (s1) -- (s2);
				\draw [arrow] (s2) -- (s3);
				\draw [arrow] (s3) -- (s4);
			\end{tikzpicture}
		\end{center}
		\caption{The diagram of the whole process of constructing the proposed filterbank.}\label{diagram}
		\label{fig:whole}
	\end{figure}
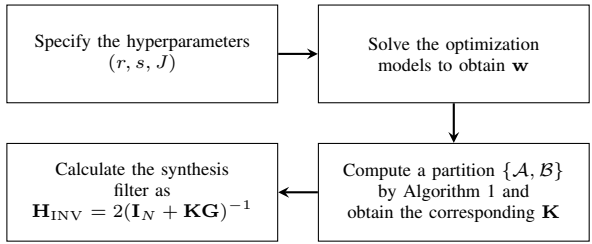

	Since $\bfG=\bfU\boldsymbol{\Gamma}\bfU^\top$, we actually need to determine $\boldsymbol{\Gamma}$ and $\bfK$. The entire process is as follows: first, we provide Theorem \ref{thm:main} which states the sufficient conditions for $\boldsymbol{\Gamma}$ and $\bfK$ to satisfy ($1^\circ$) and ($2^\circ$).
	Second, according to the theorem, we formulate optimization models in Section \ref{sec:opt} to compute the weights $\bfw$, which determines $\boldsymbol{\Gamma}$, and provide an algotithm in Section \ref{sec:partition} to partition the vertex set $\mathcal{V}$ into two disjoint subsets $\{\mcA,\mcB\}$, which gives $\bfK$.
The process of constructing the proposed filterbank is shown in Figure \ref{fig:whole}.

	In the following, let $\mathcal{I}_r=\{1,...r\}$, $\mathcal{I}_s=\{N+1-s,...,N\}$. $\bfU(\mcA,\mathcal{I}_r)$ denotes the submatrix consisting of entries of $\bfU$ with row indices in $\mcA$ and column indices in $\mathcal{I}_r$. $\bfU(\mcB,\mathcal{I}_s)$ has a similar meaning.
	
	%%----------------------------------------------------------
	\begin{thm}\label{thm:main}
		Given  $r\ge1, s\geq1$ satisfying $r+s\le N$. 
		Assume that the eigenvalues of $\bfG$ satisfy 
		\begin{equation}\label{eq:thm:main-1}
			\gamma_1=\cdots=\gamma_r=1,~~
			\gamma_{N+1-s}=\cdots=\gamma_N=-1
		\end{equation}
		and one of the following two sets of conditions:
		\begin{align}\label{eq:thm:main-2}
			\begin{split}
				&|\gamma_i|<1,~~\forall r<i<N+1-s,\\
				&\text{or}\\
				&|\gamma_i|>1,~~\forall r<i<N+1-s.
			\end{split}
		\end{align}
		Then
		\begin{align}\label{eq:annihilation}
			\begin{split}
				&(\bfI_N-\bfG)\bfu_i=\mathbf{0},~~\forall i=1,...,r,\\
				&(\bfI_N+\bfG)\bfu_i=\mathbf{0},~~\forall i=N-s+1,...,N.
			\end{split}
		\end{align}
		Furthermore, if the vertex set $\mathcal{V}=\{1,...,N\}$ can be partitioned into two disjoint subsets $\mcA,\mcB$ such that both the submatrices $\bfU(\mcA,\mathcal{I}_r)$ and $\bfU(\mcB,\mathcal{I}_s)$ are of full column rank, then $\bfI_N+\bfK\bfG$ is invertible, where $\bfK$ is a diagonal matrix satisfying 
		\begin{align}\label{eq:K}
			\bfK(i,i)=\begin{cases}
				1,&i\in \mcA,\\
				-1,&i \in \mcB.
			\end{cases}
		\end{align}
	\end{thm}
	%%----------------------------------------------------------
	\prf
	1) By $\bfG=\bfU\boldsymbol{\Gamma}\bfU^\top$, it is easy to prove \eqref{eq:annihilation}.
	
	2) Suppose $\bfx\in\br^N$ lies in the null space of $\bfI_N+\bfK\bfG$, i.e., $(\bfI_N+\bfK\bfU\boldsymbol{\Gamma}\bfU^\top)\bfx=0$. 
	Let $\bfy=\bfU^\top\bfx$, then
	\begin{align}
		\begin{split}
			0&=\|\bfx\|_2^2-\|\bfK\bfU\boldsymbol{\Gamma}\bfU^\top\bfx\|_2^2\\
			&=\|\bfy\|_2^2-\|\boldsymbol{\Gamma}\bfy\|_2^2=\sum_{i=1}^{N}(1-\gamma_i^2)y_i^2.
		\end{split}
	\end{align}
	The second equality holds because the orthogonal transformation preserves the $2$-norm of a vector and both $\bfK,\bfU$ are orthogonal matrices.
	
	Since $\gamma_i^2<1$ or $\gamma_i^2>1$ for all $r<i<N+1-s$, all corresponding $y_i$ are $0$. Thus, 
	\begin{align}\label{eq:pf_thm_1}
		\bfx=\bfU\bfy=\sum_{i=1}^{r}y_i\bfu_i+\sum_{i=N+1-s}^{N}y_i\bfu_i,
	\end{align}
	and 
	\begin{align}\label{eq:pf_thm_2}
		\bfG\bfx=\bfU\boldsymbol{\Gamma}\bfy=\sum_{i=1}^{r}y_i\bfu_i-\sum_{i=N+1-s}^{N}y_i\bfu_i.
	\end{align}
	Denote $\bfx_1=\sum_{i=1}^{r}y_i\bfu_i$ and $\bfx_{-1}=\sum_{i=N+1-s}^{N}y_i\bfu_i$. Combining \eqref{eq:pf_thm_1} and \eqref{eq:pf_thm_2} with $(\bfI_N+\bfK\bfG)\bfx=\bf0$ gives
	\begin{align}\label{eq:pf_thm_3}
		\bfx_1+\bfx_{-1}+\bfK\bfx_1-\bfK\bfx_{-1}=\bf0.
	\end{align}
	Premultiplying $\bfK$ on both sides of \eqref{eq:pf_thm_3} gives
	\begin{align}\label{eq:pf_thm_4}
		\bfx_1-\bfx_{-1}+\bfK\bfx_1+\bfK\bfx_{-1}=\bf0.
	\end{align}
	Calculating the sum and difference of \eqref{eq:pf_thm_3} and \eqref{eq:pf_thm_4} shows that
	\begin{align}
		(\bfI_N+\bfK)\bfx_1=(\bfI_N-\bfK)\bfx_{-1}=\bf0.
	\end{align}
	According to the definition of $\bfK$, there must hold
	\begin{align}
		\bfx_1(i)=0,~\forall i\in \mcA,~~~~\bfx_{-1}(i)=0,~\forall i\in \mcB.
	\end{align}
	Since $\bfx_1=\sum_{i=1}^{r}y_i\bfu_i$ and $\bfU(\mcA,\mathcal{I}_r)$ has full column rank, there holds $y_1=\cdots=y_r=0$. Similarly, $y_{N+1-s}=\cdots=y_N=0$. Therefore, $\bfx=\bfx_1+\bfx_{-1}=\bf0$, and $\bfI_N+\bfK\bfG$ is invertible. $\bbox$
	
	\textit{Remark}: Note that if both $r$ and $s$ are too large, there may not exist a partition $\{\mcA,\mcB\}$ of vertices such that both $\bfU(\mcA,\mathcal{I}_r)$ and $\bfU(\mcB,\mathcal{I}_s)$ are of full column rank. But in practice, we usually set $r$ and $s$ to be numbers much smaller than $N$, in which case finding such a partition is generally not difficult and even full of options.

	Next, consider a special case where the intrinsic graph $\mathcal{G}_b=\{\mathcal{V}_b,\mathcal{E}_b,\bfA_b\}$ is bipartite, that is, 
	the vertex set $\mathcal{V}_b$ can be partitioned into two disjoint subsets $\mcA,\mcB$ (which are called two parts of $\mathcal{G}_b$) such that connections  exist only between $\mcA$ and $\mcB$. Then a natural sampling pattern is to keep one of the two parts in the lowpass channel and the other in the highpass channel \cite{narang2012perfect,narang2013compact}. We will provide sufficient conditions for the filterbank to be perfectly reconstructed under this sampling pattern.
	
We first introduce some notations. Suppose $\mathcal{G}_b$ is connected. Let $\bfA_b^\s$ be the normalized adjacency matrix of $\mathcal{G}_b$ whose eigendecomposition is  $\bfA_b^\s=\bfU_b\diag(\xi^b_1,...,\xi_N^b)\bfU_b^\top$. The eigenvalues $\{\xi_i^b\}_{i=1}^N$ are assumed to be in \textbf{descending} order. Similarly, define  $\bfG_b=\sum_{l=1}^{J}w_l(\bfA_b^\s)^{l-1}$. It can also be written as $\bfG_b=\bfU_b\diag(\gamma_1^b,...,\gamma_N^b)\bfU_b^\top$, where $\gamma_i^b=\sum_{l=1}^{J}w_l(\xi_i^b)^{l-1}$, $\forall i=1,...,N$.
	%%----------------------------------------------------------
	\begin{prop}\label{prop:bipart}
		If $1\leq r,s\leq\frac{1}{2}\rank(\bfA_b^\s)$, then both $\bfU_b(\mcA,\mathcal{I}_r)$ and $\bfU_b(\mcB,\mathcal{I}_s)$ 
		are full column rank. Define $\bfK_b$ as a diagonal matrix with 
		\begin{align}
			\bfK_b(i,i)=\begin{cases}
				1,&i\in \mcA,\\
				-1,&i \in \mcB.
			\end{cases}
		\end{align}
		If the eigenvalues $\{\gamma_i^b\}_{i=1}^N$ of $\bfG_b$ satisfy the conditions 
		\eqref{eq:thm:main-1} and \eqref{eq:thm:main-2} 
		in Theorem \ref{thm:main}, then $\bfI_N+\bfK_b\bfG_b$ is invertible.
	\end{prop}
	%%----------------------------------------------------------
	\prf Since $\mathcal{G}_b$ is bipartite, we can label the vertices so that
	\begin{align*}
		\bfA_b^\s=\begin{pmatrix}
			\bf0 & \mathbf{R}\\
			\mathbf{R}^\top & \bf0
		\end{pmatrix},
	\end{align*}
	where $\mathbf{R}\in\br^{|\mcA|\times|\mcB|}$. 
	Denote $\bfv_i$ the $i$-th column of $\bfU_b$ and write $\bfv_i$ as $\bfv_i=[\bfv_{i\mcA}^\top,\bfv_{i\mcB}^\top]^\top$, where $\bfv_{i\mcA}$  and $\bfv_{i\mcB}$ 
	are the subvectors of $\bfv_i$ whose indices are respectively in $\mcA$ and $\mcB$.
	
	Since $\mathcal{G}_b$ is bipartite, it is known that if $[\bfv_{i\mcA}^\top,\bfv_{i\mcB}^\top]^\top$ is an eigenvector of $\bfA_b^\s$ associated with eigenvalue $\xi_i^b$, then $[\bfv_{i\mcA}^\top,-\bfv_{i\mcB}^\top]^\top$ is an eigenvector of $\bfA_b^\s$ associated with eigenvalue $-\xi_i^b$ \cite{chung1997spectral}. 
	Suppose $\{\xi^b_i\}^N_{i=1}$ has $p$ positive terms, then it also has $p$ negative terms. Hence, $\rank(\bfA_b^\s)=2p$.

	For any $1\le i\le N$ satisfying $\xi^b_i\neq 0$, we have
	\begin{align}
		\bfA^\s_b\begin{bmatrix}
			\bfv_{i\mcA}\\
			\bfv_{i\mcB}
		\end{bmatrix}=\xi^b_i\begin{bmatrix}
			\bfv_{i\mcA}\\
			\bfv_{i\mcB}
		\end{bmatrix},~~
		\bfA^\s_b\begin{bmatrix}
			\bfv_{i\mcA}\\
			-\bfv_{i\mcB}
		\end{bmatrix}=\xi^b_i\begin{bmatrix}
			-\bfv_{i\mcA}\\
			\bfv_{i\mcB}
		\end{bmatrix}.
	\end{align}
	Adding or subtracting these two equations gives
	\begin{align}
		\bfA^\s_b\begin{bmatrix}
			\bfv_{i\mcA}\\
			\mathbf{0}
		\end{bmatrix}=\xi^b_i\begin{bmatrix}
			\mathbf{0}\\
			\bfv_{i\mcB}
		\end{bmatrix},~~
		\bfA^\s_b\begin{bmatrix}
			\mathbf{0}\\
			-\bfv_{i\mcB}
		\end{bmatrix}=\xi^b_i\begin{bmatrix}
			\bfv_{i\mcA}\\
			\mathbf{0}
		\end{bmatrix},
	\end{align}
	which implies that $\bfv_{i\mcA}=\mathbf{0}$ if and only if $\bfv_{i\mcB}=\mathbf{0}$. Since 
	$\|\bfv_{i\mcA}\|_2^2+\|\bfv_{i\mcB}\|_2^2=\|\bfv_i\|_2^2=1$, we conclude that $\bfv_{i\mcA}$ and $\bfv_{i\mcB}$ are both non-zero.
	\par
	
	Now we turn to prove that $\bfU_b(\mcA,\mathcal{I}_r)$ has full column rank. For any $i<j\le r\leq p$, since $\{\xi_i^b\}_{i=1}^N$ are in descending order, we have $\xi^b_i,\xi^b_j>0$. Therefore, as discussed above, both $\bfv_{i\mcA}$ and $\bfv_{j\mcA}$ are non-zero vectors. Considering $\xi^b_i>0$, we know that $-\xi^b_i<0$ is also an eigenvalue of $\bfA_b^\s$ whose associated eigenvector is $\bfv_{-i}=[\bfv_{i\mcA}^\top,-\bfv_{i\mcB}^\top]^\top$. Thus,
	\begin{align}
		\begin{split}
			&\bfv_i^\top\bfv_j=\bfv_{i\mcA}^\top\bfv_{j\mcA}+\bfv_{i\mcB}^\top\bfv_{j\mcB}=\bf0,\\
			&\bfv_{-i}^\top\bfv_j=\bfv_{i\mcA}^\top\bfv_{j\mcA}-\bfv_{i\mcB}^\top\bfv_{j\mcB}=\bf0,
		\end{split}
	\end{align}
	which implies that $\bfv_{i\mcA}^\top\bfv_{j\mcA}=\bf0$. Consequently, $\bfU_b(\mcA,\mathcal{I}_r)$ has full column rank.
	\par
	Similarly, we can show that $\bfU_b(\mcB,\mathcal{I}_s)$ also has full column rank.
	The invertibility of $\bfI_N+\bfK_b\bfG_b$ is a direct consequency of Theorem \ref{thm:main}. \bbox

	Proposition \ref{prop:bipart} shows that we can employ the commonly used sampling pattern when the graph is bipartite. Besides, other sampling patterns can also be chosen as long as the conditions proposed in Proposition \ref{prop:bipart} are met. This is useful when the bipartite graph has an unbalanced partition of vertices, i.e. the sizes of the two parts $|\mcA|,|\mcB|$ differ a lot in which case the natural sampling pattern may lead to a low compression ratio (keep the larger part in LP channel) or a great loss of information (keep the smaller part in LP channel). Then we can search for other sampling patterns to produce a balanced partition that satisfy the conditions.

	\subsection{Formulating the Optimization Problems}\label{sec:opt}
	By definition, $\bfG$ is determined by the weights $\bfw\in\br^J$ when the graph is given. In order to obtain the desired filter responses, optimization models will be formulated to compute $\bfw$. For example, we can minimize $\|\bfH_\lp-\bfH^{\text{ideal}}\|_2$ to make $\bfH_\lp$ approximate the ideal lowpass filter $\bfH^{\text{ideal}}$ whose response is given as:
	\begin{align}
		\bfh^{\text{ideal}}(k)=\begin{cases}
			1,&\xi_k\geq\xi_0,\\
			0,&\mbox{otherwise},
		\end{cases},~k=1,...,N
	\end{align}
	where $\xi_0\in [\xi_N,\xi_1]$ is a pre-determined threshold and $\{\xi_k\}_{k=1}^N$ are the eigenvalues of $\bfA^\s$ in descending order. 
	
	We will list the constraints of the optimization model to meet the conditions in Theorem \ref{thm:main}. Without loss of generality, assume that the eigenvalues $\{\xi_i\}_{i=1}^N$ of $\bfA^\s$ are distinct. For a fixed $J\geq2$, let $\bfC\in\br^{N\times J}$ be the  Vandermonde matrix generated by $\{\xi_i\}_{i=1}^N$, i.e.,
	\begin{align*}
		\bfC=\begin{bmatrix}
			1&\xi_1&\cdots&\xi_1^{J-1}\\
			1&\xi_2&\cdots&\xi_2^{J-1}\\
			\vdots&\vdots&\ddots&\vdots\\
			1&\xi_N&\cdots&\xi_N^{J-1}\\
		\end{bmatrix},
	\end{align*}
	recall that $\bfG=\sum_{l=1}^{J}w_l(\bfA^\s)^{l-1}=\bfU\boldsymbol{\Gamma}\bfU^\top$, then $\boldsymbol{\Gamma}=\diag(\bfC\bfw)$. Thus, the analysis filter responses are given as
	\begin{align}
		\bfh_\lp=\frac{1}{2}(\bf1+\bfC\bfw),~~\bfh_\hp=\frac{1}{2}(\bf1-\bfC\bfw).
	\end{align}
	Let $\bfC_r,\bfC_s$ and $\bfC_m$ respectively denote the submatrices formed by the first $r$ rows, the last $s$ rows and the rest rows of $\bfC$.  Consider the first set of conditions in Theorem \ref{thm:main}:
	\begin{align*}
		&\gamma_1=\cdots=\gamma_r=1,~\gamma_{N+1-s}=\cdots=\gamma_N=-1,\\
		&\gamma_i\in(-1,1),~~i=r+1,...,N-s,
	\end{align*}
and construct such a convex optimization problem:
	\begin{align}\label{opt:main}
		\begin{cases}
			\underset{\bfw\in\br^J}{\mbox{min}}&\|\bfh^{\text{ideal}}-\frac{1}{2}(\bf1+\bfC\bfw)\|_{\infty}\\
			\mbox{s.t.}&\bfC_r\bfw=\bf1_r\\
			\quad&\bfC_s\bfw=-\bf1_s\\
			\quad&|\bfC_m\bfw|<\mathbf{1}_{N-r-s}
		\end{cases}.
	\end{align}
	Note that the objective function is actually equivalent to $\|\bfH^{\text{ideal}}-\bfH_\lp\|_2$.

	When $r=s=1$, the problem \eqref{opt:main} is always feasible for any $J\geq2$, since
$$\bfw=[-\frac{\xi_N+1}{1-\xi_N},\frac{2}{1-\xi_N},0,...,0]\in\br^J$$ is in the feasible domain (note that $\xi_1=1$). While in other cases, one should pay close attention to the feasibility of the problem since we cannot guarantee that the feasible domain is non-empty for all settings  of $r,s,J$. Therefore, it is recommended to take $r=s=1$ if you do not want to test the feasibility of the problem with $r,s,J$ in other settings.
	
\begin{figure*}[htbp]
		\centering{
			\includegraphics[width=0.3\textwidth]{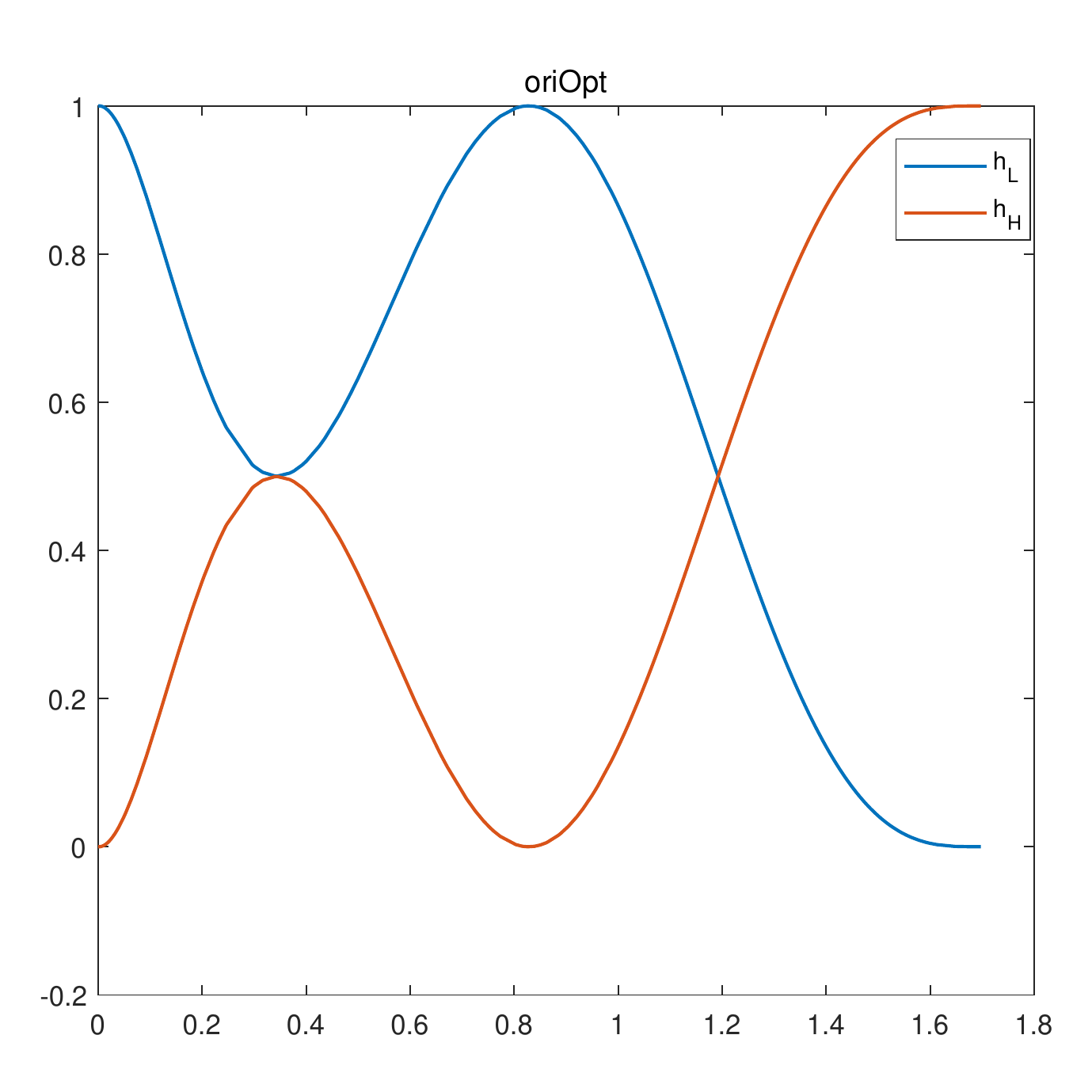}
			\includegraphics[width=0.3\textwidth]{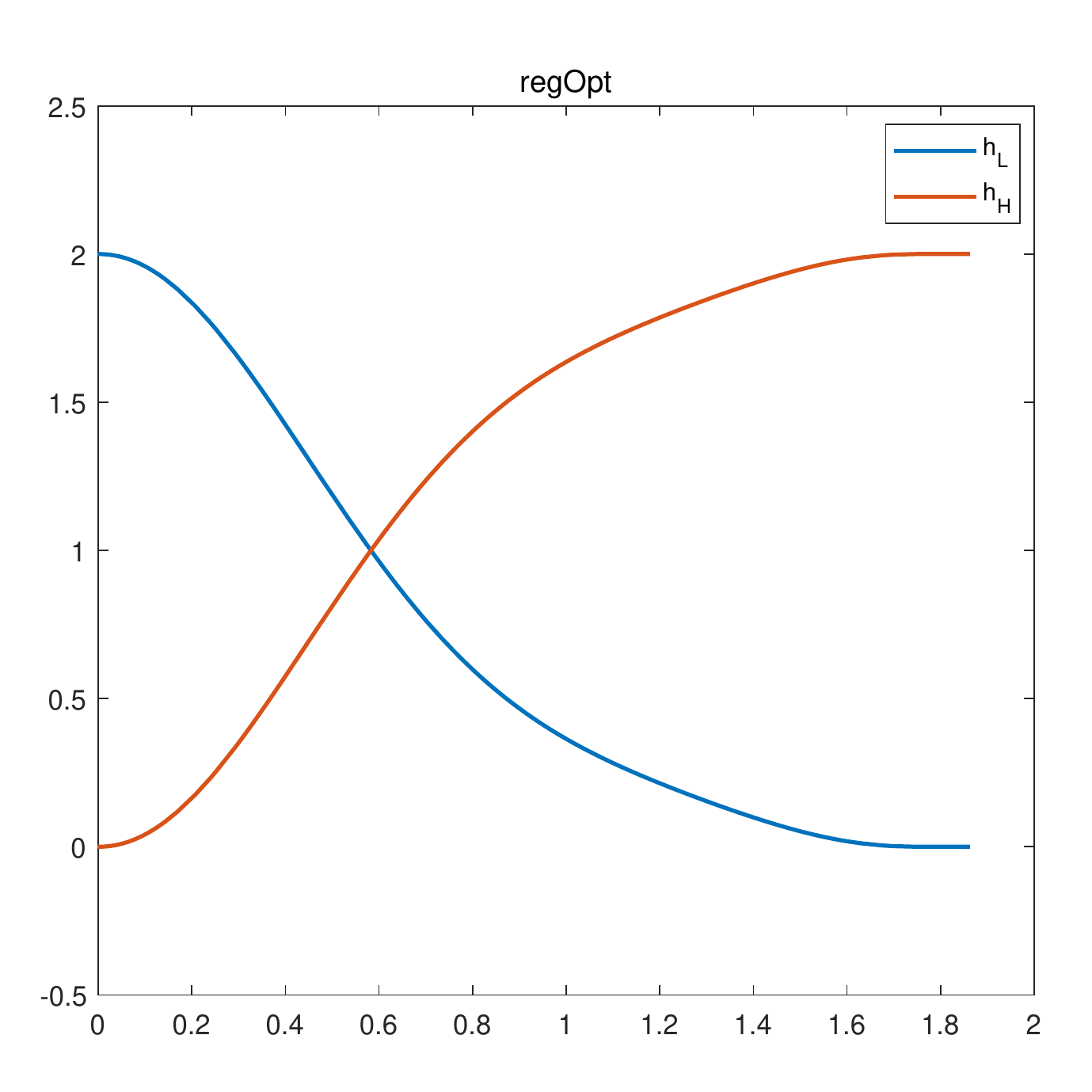}
			\includegraphics[width=0.3\textwidth]{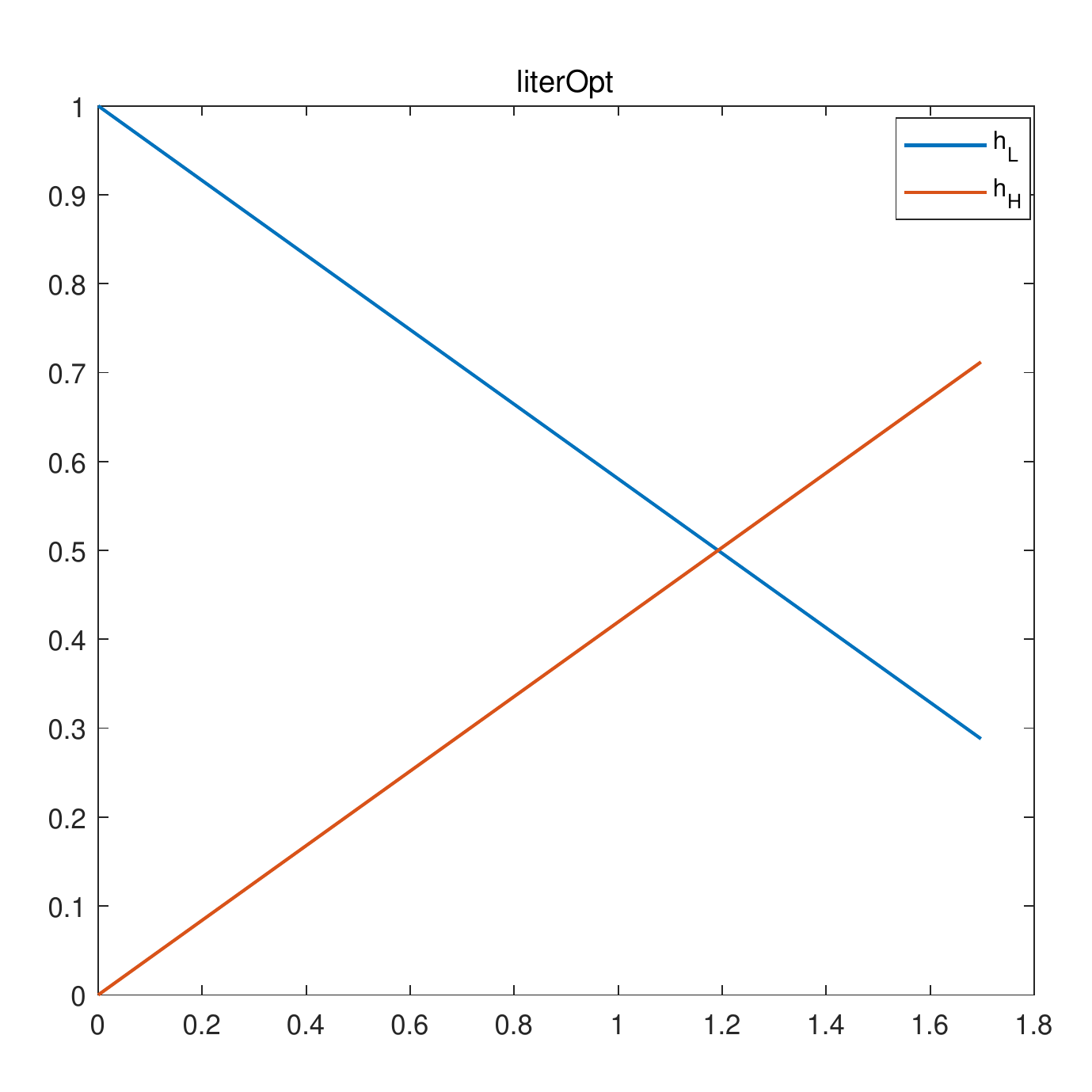}
		}
		\caption{\small 
			From left to right are the filter responses produced respectively by the three optimization models: oriOpt, regOpt and literOpt, blue for lowpass and red for highpass.
			The horizontal axis represents the eigenvalues of $\bfL^\s$, in which the
			duplicate eigenvalues have been removed before solving the optimization problems.}
		\label{fig:h_l}
	\end{figure*}

	Recall that the spline-like filters are localized in the graph domain, and the smaller $J$ is, the better the locality. However, low-order polynomials may not provide a good approximation of the ideal lowpass filter, as shown in the left  of Figure \ref{fig:h_l}, unexpected peaks and valleys may appear in the middle section of the polynomial filter response. For this rationale, we would like to add a regularization term $R(\bfh_\lp)$ to the original objective function to improve the smoothness of $\bfh_\lp$. Denote $p_\lp$ the polynomial associated with $\bfh_\lp$, i.e., 
$$p_\lp(x)=1+\sum_{l=1}^{J}w_lx^{l-1},~~~x\in[\xi_N,\xi_1].$$ 
Consider the $2$-norm of function $p_\lp$:
	\begin{align}
		\|p_\lp\|_{2}=\int_{\xi_N}^{\xi_1}\big|p'_\lp(x)\big|^2dx
		=\int_{\xi_N}^{\xi_1}\Big|\sum_{l=1}^{J-1}w_{l+1}lx^{l-1}\Big|^2dx.
	\end{align}
	Let $R(\bfh_\lp)$ be the discrete version:
	\begin{align}
		R(\bfh_\lp)=\sum_{i=1}^{N}|p'_\lp(\xi_i)|^2=\|\bfC_{0}\diag(0,...,J-1)\bfw\|_2,
	\end{align}
	where 
	\begin{align*}
		\bfC_0=\begin{bmatrix}
			0&1&\xi_1&\cdots&\xi_1^{J-2}\\
			0&1&\xi_2&\cdots&\xi_2^{J-2}\\
			\vdots&\vdots&\vdots&\ddots&\vdots\\
			0&1&\xi_N&\cdots&\xi_N^{J-2}\\
		\end{bmatrix}\in\br^{N\times J}.
	\end{align*}
Then, the regularized optimization problem is
	\begin{align}\label{opt:regularize}
		\begin{cases}
			\underset{\bfw\in\br^J}{\mbox{min}}&\|\bfh^{\text{ideal}}-\frac{1}{2}(\mathbf{1}+\bfC\bfw)\|_{\infty}+\alpha R(\bfh_\lp)\\
			\mbox{s.t.}&\bfC_r\bfw=\bf1_r\\
			\quad&\bfC_s\bfw=-\bf1_s\\
			\quad&|\bfC_m\bfw|<\mathbf{1}_{N-r-s}
		\end{cases},
	\end{align}
	where $\alpha\geq0$ is a parameter that controls the importance of the regularization term. 
	
	For simplicity, we refer to the proposed two optimization models \eqref{opt:main} and \eqref{opt:regularize}  as \textbf{oriOpt} and \textbf{regOpt} respectively, and the model \eqref{opt:cite} proposed in \cite{Ekambaram2015Spline} as \textbf{literOpt}.
	Figure \ref{fig:h_l} shows an example of the lowpass filter responses determined by these three models, where the parameters are taken to be $(r,s,J,\alpha)=(2,3,8,0.5)$, and the desired filter responses are all $\bfh^{\text{ideal}}$. In this work, we always use CVX, a package for specifying and solving convex programs \cite{cvx,gb08}, to solve the optimization problems.
	
	It is shown that regularized method outperforms the other two methods. As we have expected, oriOpt produces an oscillatory solution, which is less ideal than the smooth solution produced by regOpt. It is worth mentioning that for literOpt, we have done a lot of experiments with various values of $J$ on a lot of random sensor graphs, it always gave a linear filter response.% Figure \ref{fig:h_l}.

%------------
\subsection{Determining the Partition $\{\mcA,\mcB\}$}\label{sec:partition}
According to Theorem \ref{thm:main}, the sampling matrix $\bfK$ is determined by the partition $\{\mcA,\mcB\}$ of $\mathcal{V}$. Given a normalized adjacency matrix $\bfA^\s$, Algorithm \ref{alg:A_B} outputs a partition $\{\mcA,\mcB\}$ of $\mathcal{V}$ that makes the matrices $\bfU(\mcA,\mathcal{I}_r)$ and $\bfU(\mcB,\mathcal{I}_s)$ have full column rank. The symbol ``$\ll$'' means ``much smaller than'', and the operation $\mathcal{V}\backslash \mcA$ computes the difference between two sets.
%%--------------------------------------------------
\begin{algorithm}[htbp]
	\renewcommand{\algorithmicrequire}{\textbf{Input:}}
	\renewcommand{\algorithmicensure}{\textbf{Output:}}
	\caption{Search for $\{\mcA,\mcB\}$}
	\label{alg:A_B}
	\begin{algorithmic}[1]
		\REQUIRE Normalized adjacency matrix $\bfA^\s$, $1\leq r,s\ll N$
		\STATE Initialization: Set $\mathcal{V}=\{1,...,N\},\mcA=\emptyset,\mcB=\emptyset$, $\bfU_r=\bfU(\mathcal{V},\mathcal{I}_r),\bfU_s=(\mathcal{V},\mathcal{I}_s)$
		\STATE Compute the row echelon form of $\bfU_r$ to obtain $r$ linearly independent rows, and add their indices to $\mcA$
		\IF {$\bfU_s(\mathcal{V}\backslash \mcA,\mathcal{I}_s)$ is not full column rank}
		\STATE  Throw an error, quit and reset $r$ and $s$
		\ELSE 
		\STATE Compute the row echelon form of $\bfU_s(\mathcal{V}\backslash \mcA,\mathcal{I}_s)$ to obtain $s$ linearly independent rows, and add their indices to $\mcB$
		\STATE Partition the rest row indices into two balanced sets based on some criterion, and assign them to $\mcA$ and $\mcB$ respectively
		\ENDIF
		\ENSURE $\mcA,\mcB$
	\end{algorithmic}
\end{algorithm}
%%--------------------------------------------------

Next, we discuss how the partition $\{\mcA,\mcB\}$ will affect the approximation error of the filterbank. Let $\bfK$ be the sampling matrix defined by \eqref{eq:K}. 
If we only use the LP output $\bfy_\lp$ of the analysis stage for reconstruction, the reconstructed signal would be $\bfy'=\frac{1}{2}\bfH_{\inv}(\bfI_N+\bfK)\bfy_\lp$. Thus, the approximation error of $\bfy'$ to the original signal is $\|\bfx-\bfy'\|_2$. Since the filterbank is perfectly reconstructed, there holds $\bfy=\bfx$, where $\bfy$ is the total reconstruction defined by \eqref{eq:TCFB}. Consequently, the approximation error of the filterbank is defined as
\begin{align}\label{eq:app_er}
\begin{split}
\mbox{er}&=\|\bfy-\bfy'\|_2\\
&=\|\frac{1}{4}\bfH_{\inv}(\bfI_N-\bfK)(\bfI_N-\bfG)\bfx\|_2\\
&\leq\frac{1}{4}\|\bfH_{\inv}\|_2\|(\bfI_N-\bfK)\|_2\|(\bfI_N-\bfG)\|_2\|\bfx\|_2\\
&=2\|(\bfI_N+\bfK\bfG)^{-1}\|_2\|\bfx\|_2\\
&=\frac{2}{\sigma_{\min}(\bfI_N+\bfK\bfG)}\|\bfx\|_2,
\end{split}
\end{align}
where $\sigma_{\min}$ represents the smallest singular value of a matrix. Here we exploit the facts that $\bfH_{\inv}=2(\bfI_N+\bfK\bfG)^{-1}$ and $\|(\bfI_N-\bfK)\|_2=\|(\bfI_N-\bfG)\|_2=2$. Although $\sigma_{\min}(\bfI_N+\bfK\bfG)\neq0$ can be guaranteed by Theorem \ref{thm:main}, a small value may also lead to poor approximation. Considering that $r,s$ are usually set to be small, the method of partitioning the rest vertices in Step $7$ of Algorithm \ref{alg:A_B} is the major factor affecting the value of $\sigma_{\min}(\bfI_N+\bfK\bfG)$.

We conduct experiments to compare two strategies.
One is to partition the rest vertices according to the polarity of entries of $\bfu_N$ \cite{Shuman2016Amultiscale}: if $\bfu_N(i)<0$ then $i$ is added to $\mcB$; otherwise it
is added to $\mcA$. Another strategy is a random method which randomly partitions the rest vertices into two balanced sets $\mcA$ and $\mcB$. Other strategies can also be employed as needed.

\begin{figure}[htbp]
	\centering{
	\includegraphics[width=0.24\textwidth]{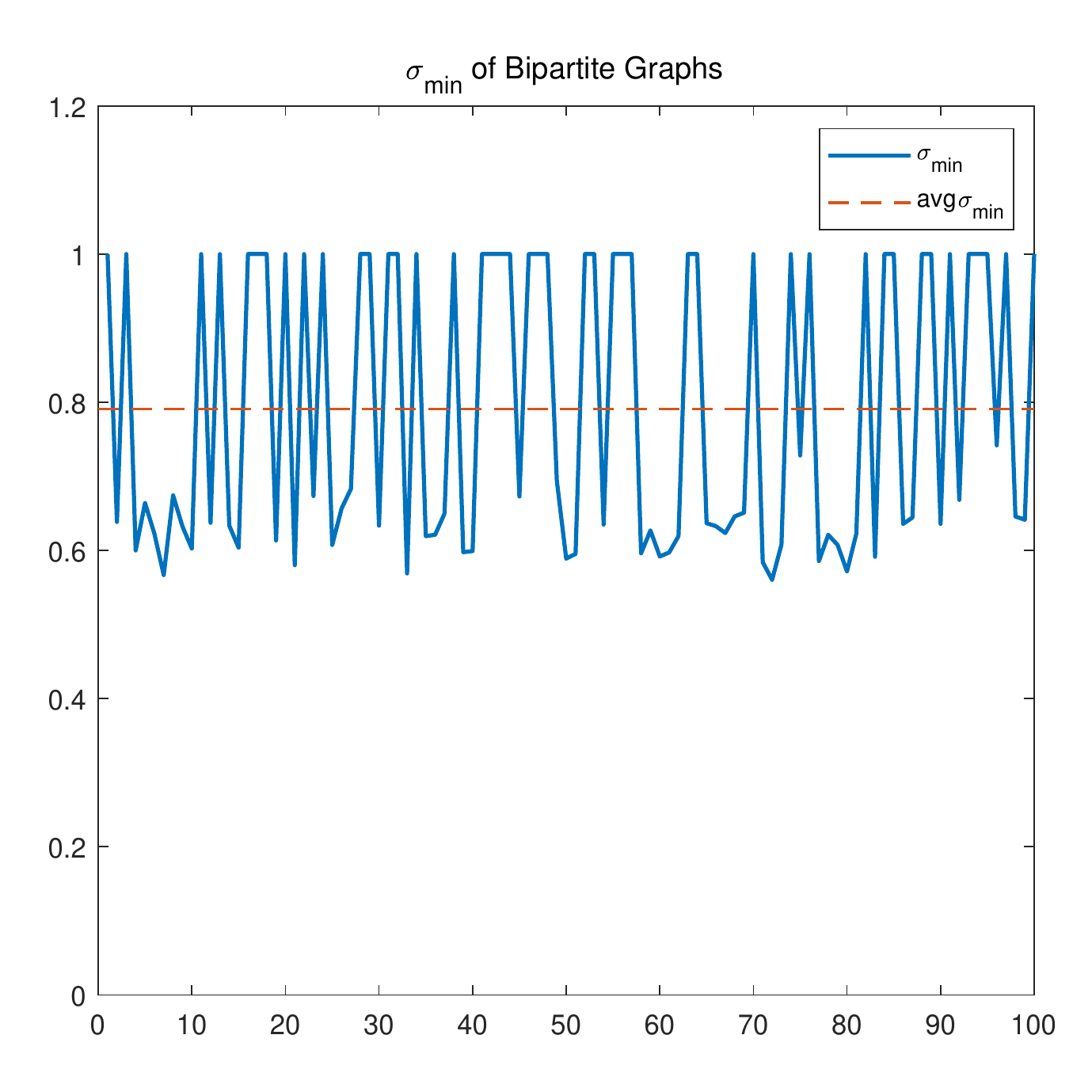}
	\includegraphics[width=0.24\textwidth]{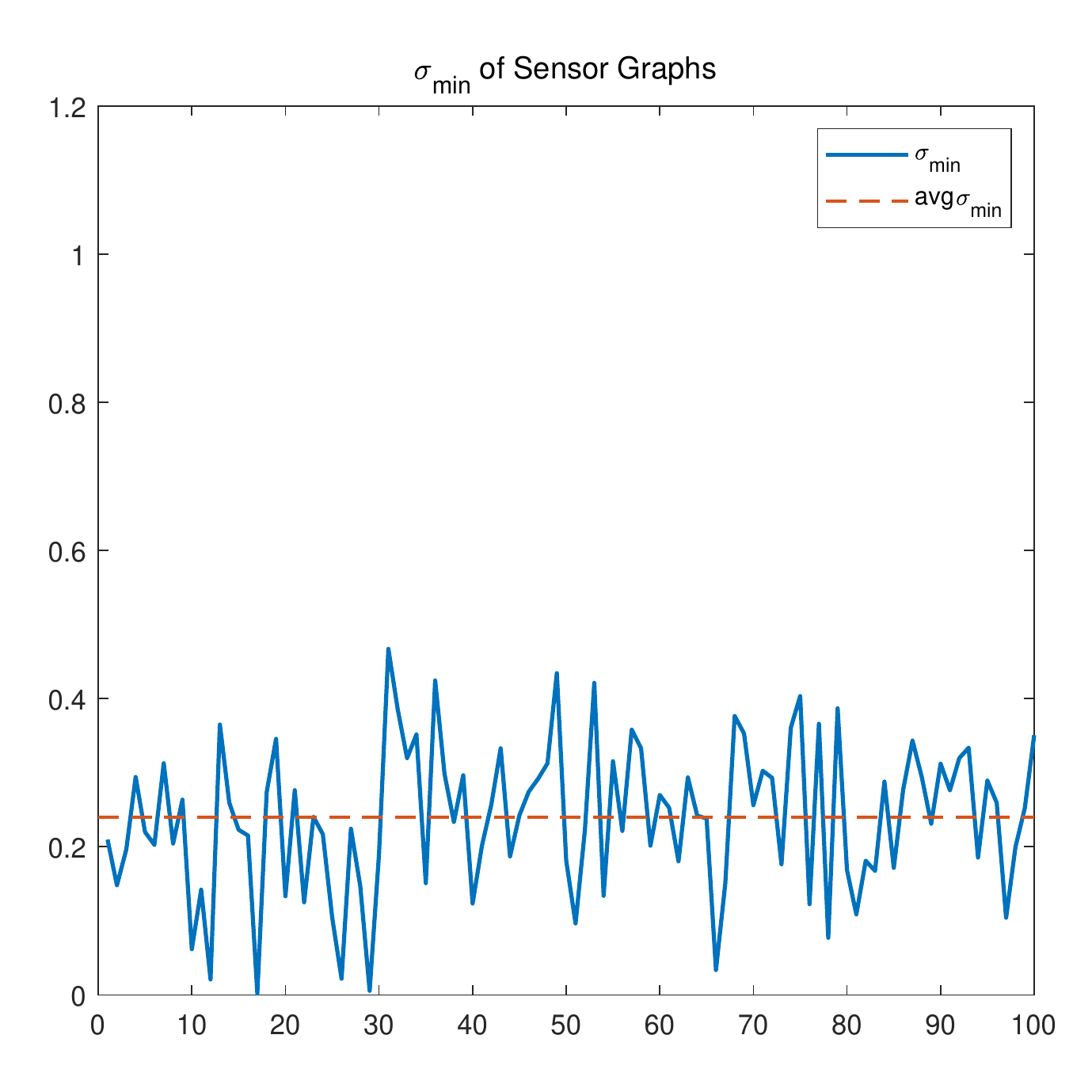}
}
\caption{\small The smallest singular values $\sigma_{\min}(\bfI_N+\bfK\bfG)$ on random graphs using the first strategy and their average $\mbox{avg} \sigma_{\min}$. Left: $\sigma_{\min}$ of random bipartite graphs; Right: $\sigma_{\min}$ of random sensor graphs.}\label{fig:min_singular_un}
\end{figure}

\begin{figure}[htbp]
	\centering{
		\includegraphics[width=0.24\textwidth]{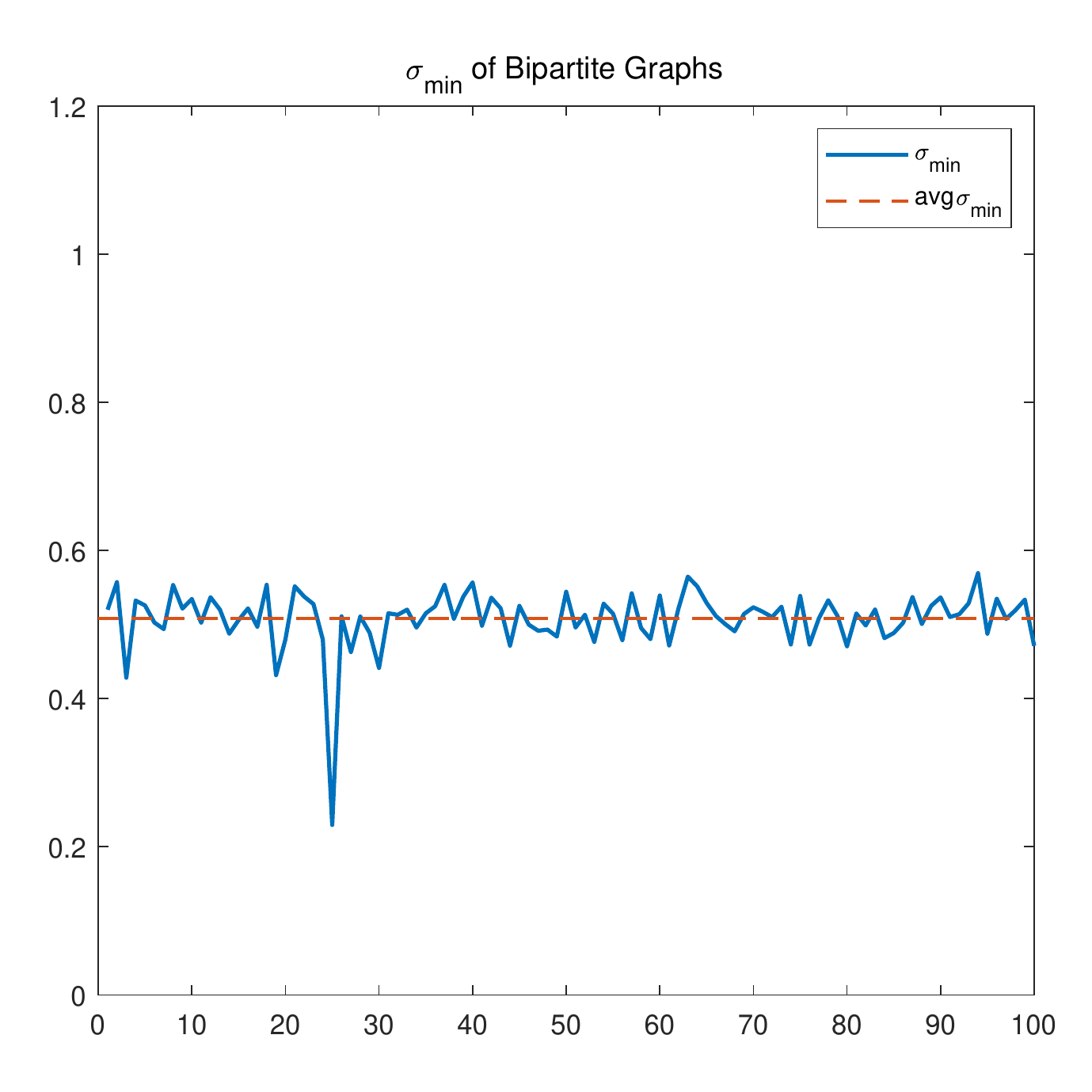}
		\includegraphics[width=0.24\textwidth]{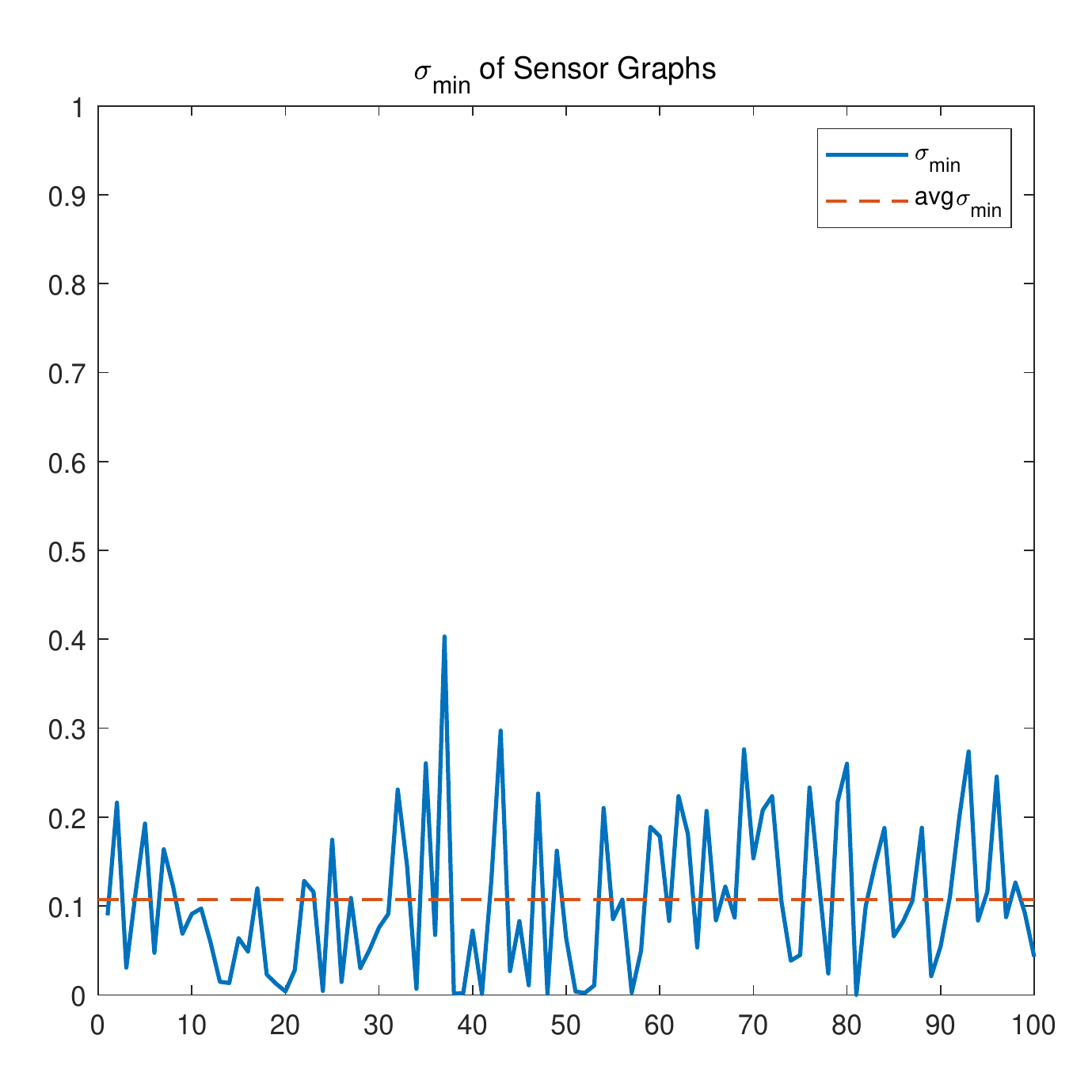}
	}
	\caption{\small The smallest singular values $\sigma_{\min}(\bfI_N+\bfK\bfG)$ on random graphs using the second strategy.}\label{fig:min_singular_ran}
\end{figure}

Experiments are performed on $100$ randomly generated bipartite graphs and $100$ random sensor graphs respectively. All graphs have $100$ vertices, and each bipartite graph has two parts of size $(20,80)$.
We solve the regOpt with $(r,s,J,\alpha)=(1,1,3,0.5)$ to obtain the weights $\bfw$ and thus $\bfG$. Figure \ref{fig:min_singular_un} and Figure \ref{fig:min_singular_ran} show the smallest singular values $\sigma_{\min}(\bfI_N+\bfK\bfG)$ using the first strategy and the second strategy respectively. It shows that the first one is better. In fact, a random strategy is not reasonable, because we need to reconnect the downsampled vertices to obtain a new graph for multi-resolution analysis. 
Therefore, we want the vertices within each set of $\{\mcA,\mcB\}$ to be connected by edges with low weights. The first strategy performs better because it is closely related to the nodal domain theory. For more details, please refer to \cite{biyikoglu2007laplacian,Shuman2016Amultiscale}.
Besides, other methods such as $k$-means clustering on $\bfu_N$ \cite{Shuman2016Amultiscale}, or sovling the max-cut problem to obtain the partition can also be used \cite{Narang2010Local}. 

\begin{figure}[htbp]
	\centering{
		\includegraphics[height=5cm,width=0.45\textwidth]{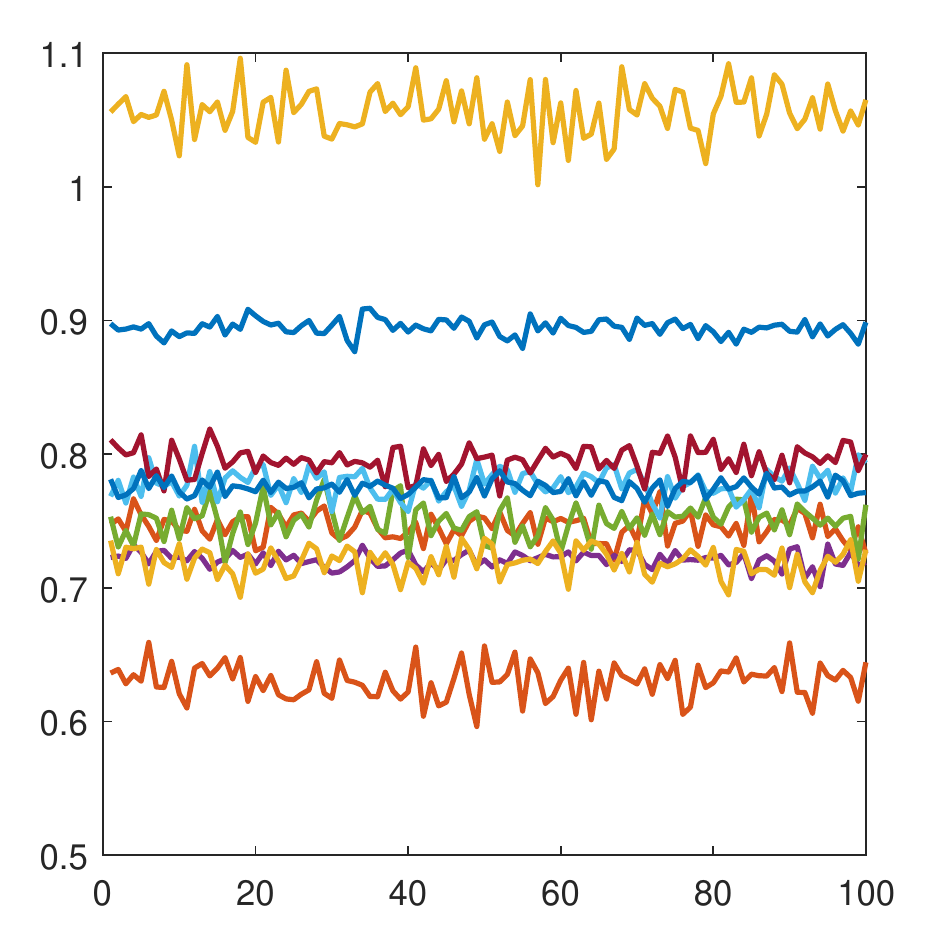}
	}
	\caption{\small The approximation errors of the proposed filterbanks on $10$ random sensor graphs. Each line represents the approximation errors resulting from the decomposition of $100$ signals on a graph. }\label{fig:approx_error}
\end{figure}

Figure \ref{fig:approx_error} shows the approximation errors of the proposed filterbanks on $10$ random sensor graphs. For each graph, we synthesize $100$ signals, each with unit norm. Only the LP output is used for reconstruction and the approximation error is computed according to the definition \eqref{eq:app_er}. We solve the regOpt with $(r,s,J,\alpha)=(1,1,3,0.5)$. The upper bounds $\frac{2}{\sigma_{\min}(\bfI_N+\bfK\bfG)}$ associated with each graph are also calculated, all of which are in the order of thousands, much greater than the approximation errors in the experiment.

\subsection{Annihilating the DC Signal}\label{subsec:DC}
In applications where the intrinsic graphs are located in the physical space, a constant signal (called DC signal) may have a physical interpretation, and the highpass filter should be able to annihilate
 the DC signal. However, the spline-like filterbanks proposed in Section \ref{subsec:main} are based on the normalized adjacency matrix $\bfA^\s$. Thus, the highpass filter $\bfH_\hp=\frac{1}{2}(\bfI_N-\bfG)$ has a zero response to $\bfu_1$, the eigenvector of $\bfL^\s$ associated with $\lambda_1=0$, which is not a constant vector unless $\mathcal{G}$ is a regular graph (i.e., all vertices have the same degree). In this case, filtering the DC signal with $\bfH_\hp$ may produce a non-zero result.
Since $\bfu_1=\bfD^{1/2}\mathbf{1}$, this problem can be addressed by pre-multiplying the input signal $\bfx$ with $\bfD^{1/2}$, and post-multiplying the filtered signal with $\bfD^{-1/2}$ \cite{narang2013compact}. Define the zero-DC analysis filters as:
\begin{equation}\label{eq:zero_filters}
	\begin{cases}
		&\bfH_\lp^0=\bfD^{-1/2}\bfH_\lp\bfD^{1/2}\\
		&\bfH_\hp^0=\bfD^{-1/2}\bfH_\hp\bfD^{1/2}
	\end{cases}.
\end{equation}
Then the whole transform of the filterbank becomes
\begin{align}
	\begin{split}
		\bfy&=\bfH_{\inv}^0\big[\frac{1}{2}(\bfI_N+\bfK)\bfH_\lp^0+\frac{1}{2}(\bfI_N-\bfK)\bfH_\hp^0\big]\bfx\\
		&=\bfH_{\inv}^0\big[\frac{1}{2}(\bfI_N+\bfK)\bfD^{-1/2}\bfH_\lp\bfD^{1/2}\\
		&+\frac{1}{2}(\bfI_N-\bfK)\bfD^{-1/2}\bfH_\hp\bfD^{1/2}\big]\bfx\\
		&=\bfH_{\inv}^0\big[\frac{1}{2}(\bfI_N+\bfK\bfD^{-1/2}\bfG\bfD^{1/2})\big]\bfx,
	\end{split}
\end{align}
where $\bfH_{\inv}^0$ represents the synthesis filter. Since $\bfD$ and $\bfK$ are commutative,
$\bfI_N+\bfK\bfD^{-1/2}\bfG\bfD^{1/2}$ is invertible iff $\bfI_N+\bfK\bfG$ is invertible. Therefore, as long as $\bfK$ and $\bfG$ satisfy the conditions proposed in Theorem \ref{thm:main}, the synthesis filter exists and is given as 
\begin{align}
	\bfH^0_{\inv}=2\bfD^{-1/2}(\bfI_N+\bfK\bfG)^{-1}\bfD^{1/2}.
\end{align}

\section{Experiments}\label{sec:expe}
In this section, we will evaluate the performance of the proposed filterbanks and compare them with related works. All experiments are done with Matlab and the GSP toolbox \cite{perraudin2014gspbox}.

First, we specify the hyperparameters $(r,s,J,\alpha)$ and solve the optimization problems to obtain the weights $\bfw$ and thus $\bfG$. Second, implement Algorithm \ref{alg:A_B} to compute a partition ${\mcA,\mcB}$ according to the polarity of the entries of $\bfu_N$, then construct the sampling matrix $\bfK$ by \eqref{eq:K}. We will employ the zero-DC filters $\bfH_\lp^0,\bfH_\hp^0,\bfH_{\inv}^0$ defined in Section \ref{subsec:DC} to form the proposed filterbanks. Multi-resolution analysis will be performed on the graph signals, thus, after downsampling, the Kron reduction scheme \cite{Dorfler2012Kron}
 is used to reconnect the vertices in $\mcA$ to produce a reduced graph, further decomposition will be recursively performed on the lowpass channel.
\begin{figure}
	\centering{
		\includegraphics[width=0.48\textwidth]{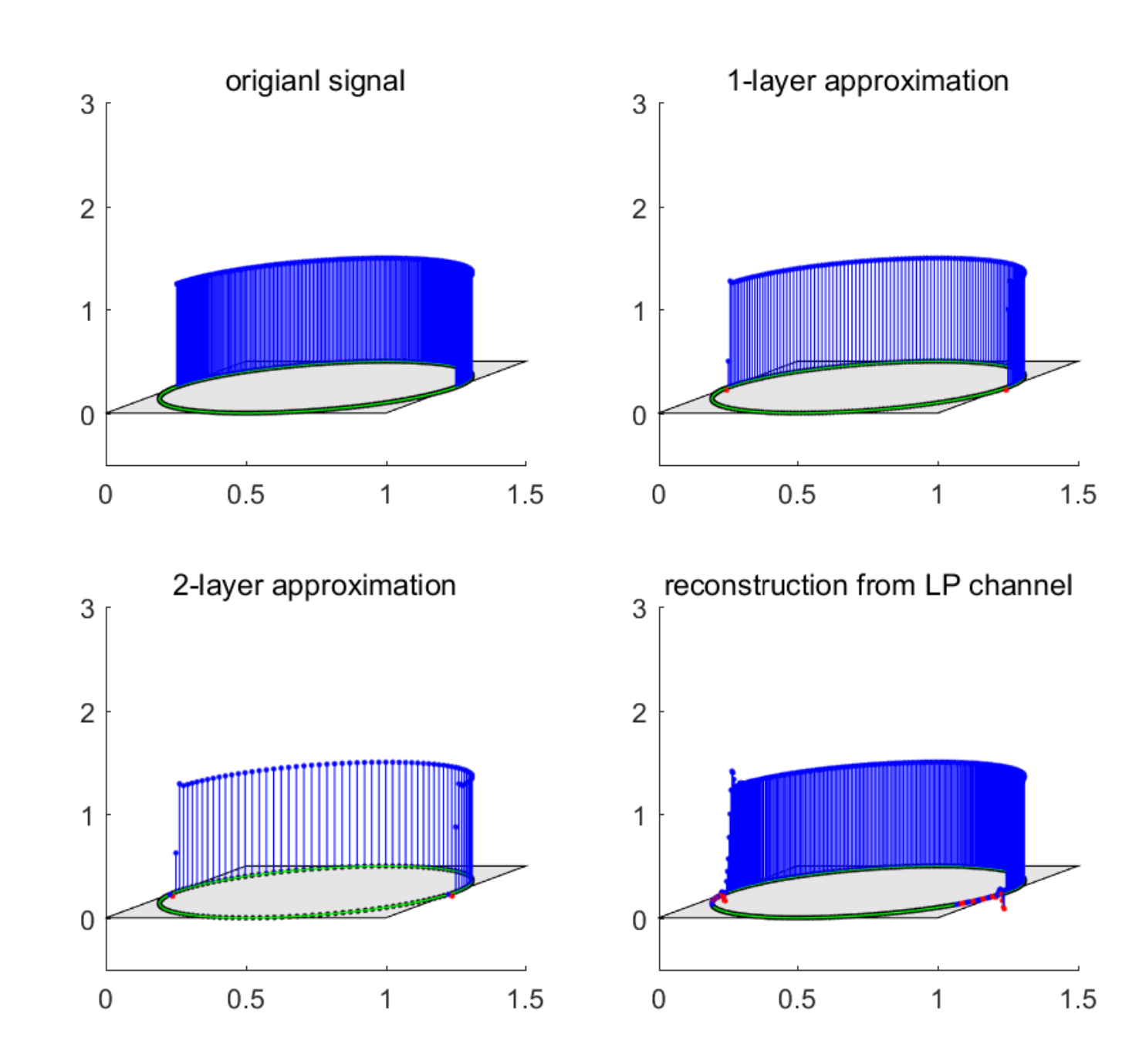}	
	}
	\caption{\small Multi-resolution analysis of the graph signal located on a bipartite ring graph. Left Top: the original signal; Right Top: the LP output in the $1$st layer; Left Bottom: the LP output in the $2$nd layer; Right Bottom: the reconstructed signal using only the LP output from the $2$nd layer decomposition.}\label{fig:locality_bi}
\end{figure}
\begin{figure}
	\centering{
		\includegraphics[width=0.48\textwidth]{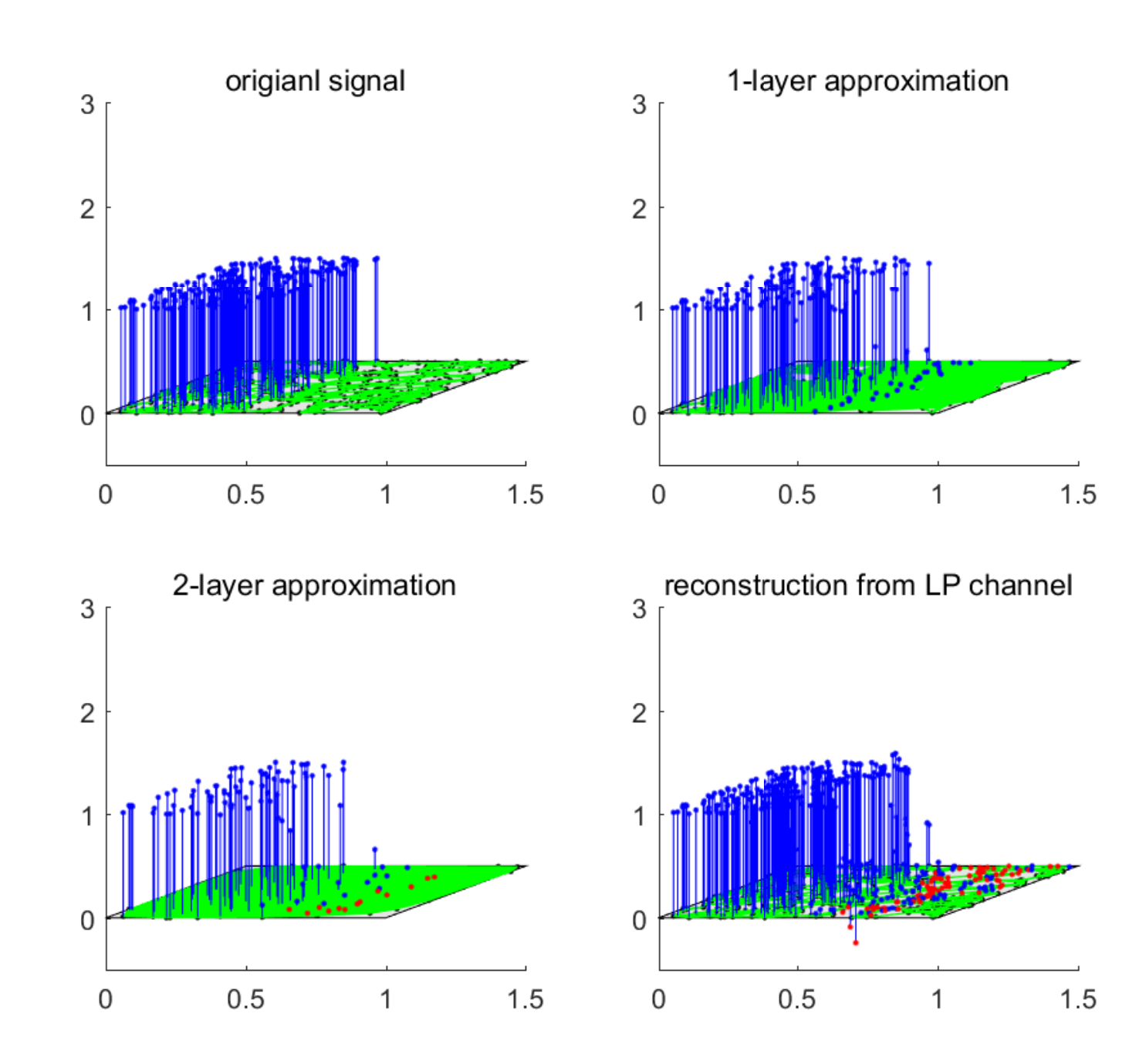}	
	}
	\caption{\small Multi-resolution analysis of the graph signal located on a sensor graph. Left Top: the original signal; Right Top: the LP output in the $1$st layer; Left Bottom: the LP output in the $2$nd layer; Right Bottom: the reconstructed signal using only the LP output from the $2$nd layer decomposition.}\label{fig:locality_se}
\end{figure}

\subsection{Locality of the Proposed Filterbanks}\label{sec:localExpe}
We use GSP toolbox to generate a bipartite ring graph with $N=512$ vertices. The corresponding graph signal $\bfx$ is  piecewise constant, i.e., the first half of $\bfx$ are all ones, and the second half are all zeros, as shown in the left top of Figure \ref{fig:locality_bi}. We solve the regOpt \eqref{opt:regularize} with $(r,s,J,\alpha)=(1,1,4,1)$. Figure \ref{fig:locality_bi} shows the LP outputs in each layer decomposition and the reconstructed signal using only the LP output of the last layer. It can be seen that there is no significant Gibbs effect near the discontinuity points of the LP outputs, indicating that the analysis filters are well localized in the graph domain.

We also compute the relative error of each LP output and the reconstructed signal. Let $\bfy_i$ be the LP output of $i$-th layer and $\bfx_i$ be the corresponding ideal output, i.e., $\bfx_i$ is a subvector of $\bfx$ whose element indices are in the downsampled subset $\mcA$ of $i$-th layer. The relative error of $\bfy_i$ with respect to (w.r.t.) $\bfx_i$ is defined as $e_i=\frac{\|\bfy_i-\bfx_i\|_2}{\|\bfx_i\|_2}$. Let $\bfy$ be the reconstructed signal, then the relative error of $\bfy$ w.r.t. $\bfx$ is $e=\frac{\|\bfy-\bfx\|_2}{\|\bfx\|_2}$. In the experiment, we get 
\[
e_1=0.032,~~e_2=0.067,~~e=0.063.
\]

Besides, the same experiment is conducted on a random sensor graph with $512$ vertices, which is not bipartite. The results are shown in Figure \ref{fig:locality_se} and the relative errors are
\[
e_1=0.057,~~e_2=0.081,~~e=0.166.
\]

\subsection{Comparison with Related Work}
In this section, we perform MRA on graph signals to compare the proposed model with related works in terms of approximation ability of the coarsened signals and denoising ability. The related shemes are literOpt \cite{Ekambaram2015Spline} and two other state-of-the-art spline-like graph filterbanks: MSGFB \cite{miraki2021modified} and SGFBSS \cite{miraki2021spline}. MSGFB is an improved model of literOpt which relaxes the constraints of the optimization problem \eqref{opt:cite} for better solution. Unlike regOpt, literOpt and MSGFB, which sample in the vertex domain, SGFBSS adopts the spectral domain sampling method proposed in \cite{tanaka2018spectral}.
\begin{figure}
	\centering{
		\includegraphics[width=0.48\textwidth]{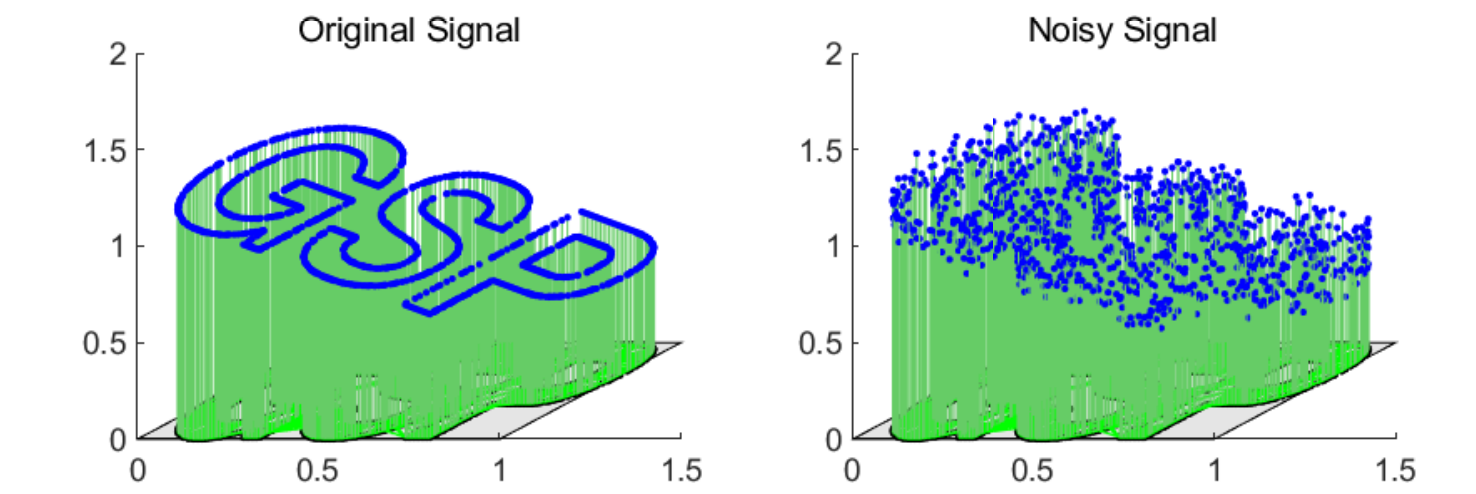}	
	}
	\caption{\small Left: the original signal; Right: the noisy signal.}\label{fig:logo_ori}
\end{figure}

\subsubsection{Approximation}
We generate the gspLogo graph with $N=1130$ and synthesize a signal $\bfx^0$ which is a linear function of the $x$-coordinates of the vertices. Then $\bfx^0$ is contaminated with Gaussian noise of zero mean and $1/16$ standard deviation to produce a noisy signal $\bfx$, as shown in Figure \ref{fig:logo_ori}. A $1$-layer decomposition is performed on the graph signal, where the hyperparameters are specified as $(r,s,J,\alpha)=(1,4,5,0.01)$ for regOpt \eqref{opt:regularize} and $J=5$ for literOpt, MSGFB and SGFBSS.

\begin{figure}[h]
	\centering{
		\includegraphics[width=0.48\textwidth]{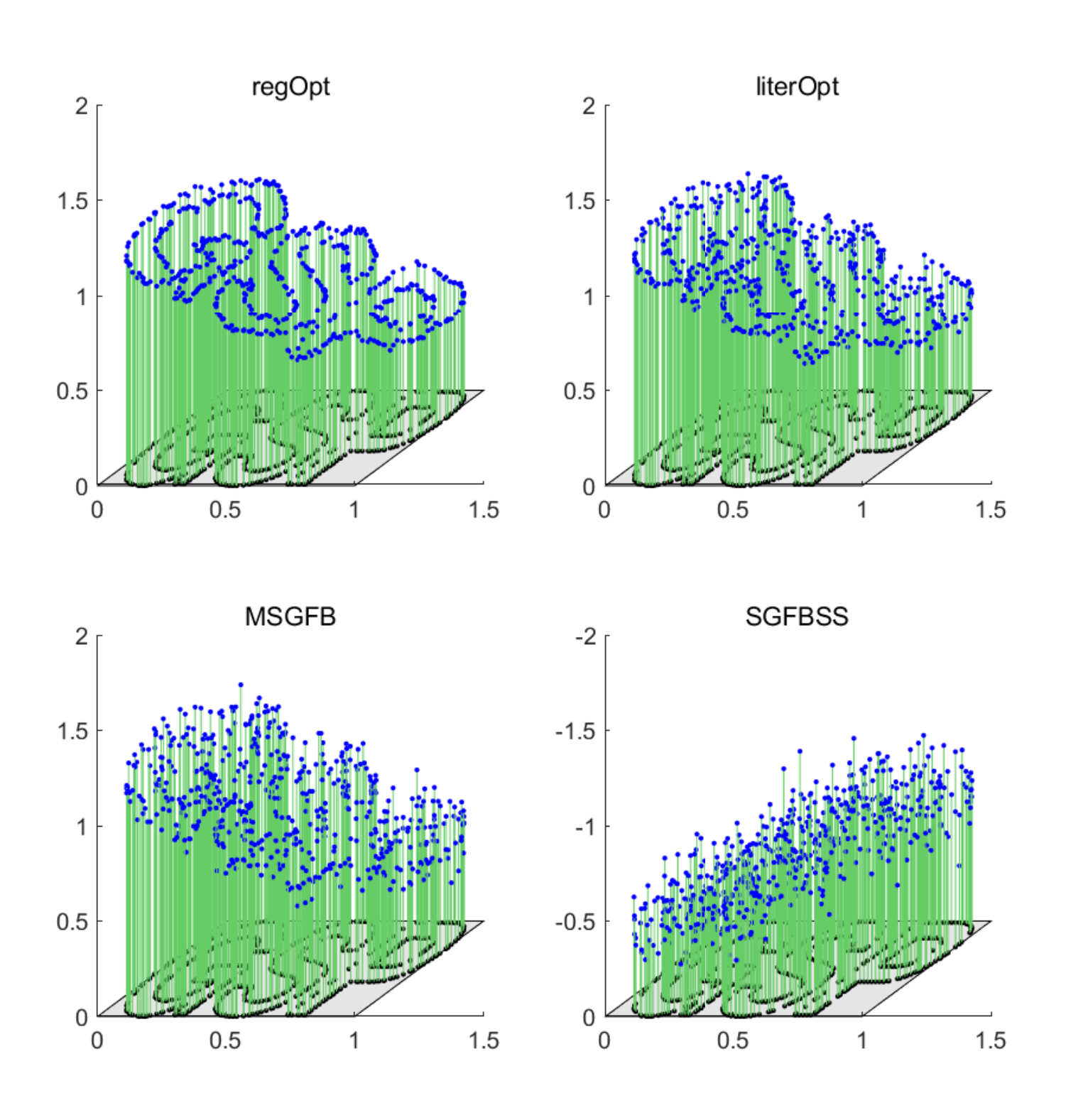}	
	}
	\caption{\small The LP outputs of each model. }\label{fig:logoMRA}
\end{figure}

In each layer decomposition of MRA, the LP output and the corresponding reduced graph serve as a coarser approximation of the original signal and graph. 
Figure \ref{fig:logoMRA} depicts the LP output of each model. It can be seen that the proposed filterbank outperforms the others. We also compute the corresponding relative errors. Let $\bfy_i$ and $\bfx_i$ have the same definitions as in Section \ref{sec:localExpe}. Then the relative errors of $\bfy_i$ w.r.t. $\bfx_i$ are $0.02, 0.04, 0.5, 1.67$ for regOpt, literOpt, MSGFB and SGFBSS, respectively.
\par

To be mentioned, the model SGFBSS cannot preserve signal values in the vertex domain due to the spectral sampling scheme, which makes the LP output differ a lot from the original signal, as shown in the right bottom image of Figure \ref{fig:logoMRA}.

\subsubsection{Denoising}
\begin{figure}
	\centering{
		\includegraphics[width=0.48\textwidth]{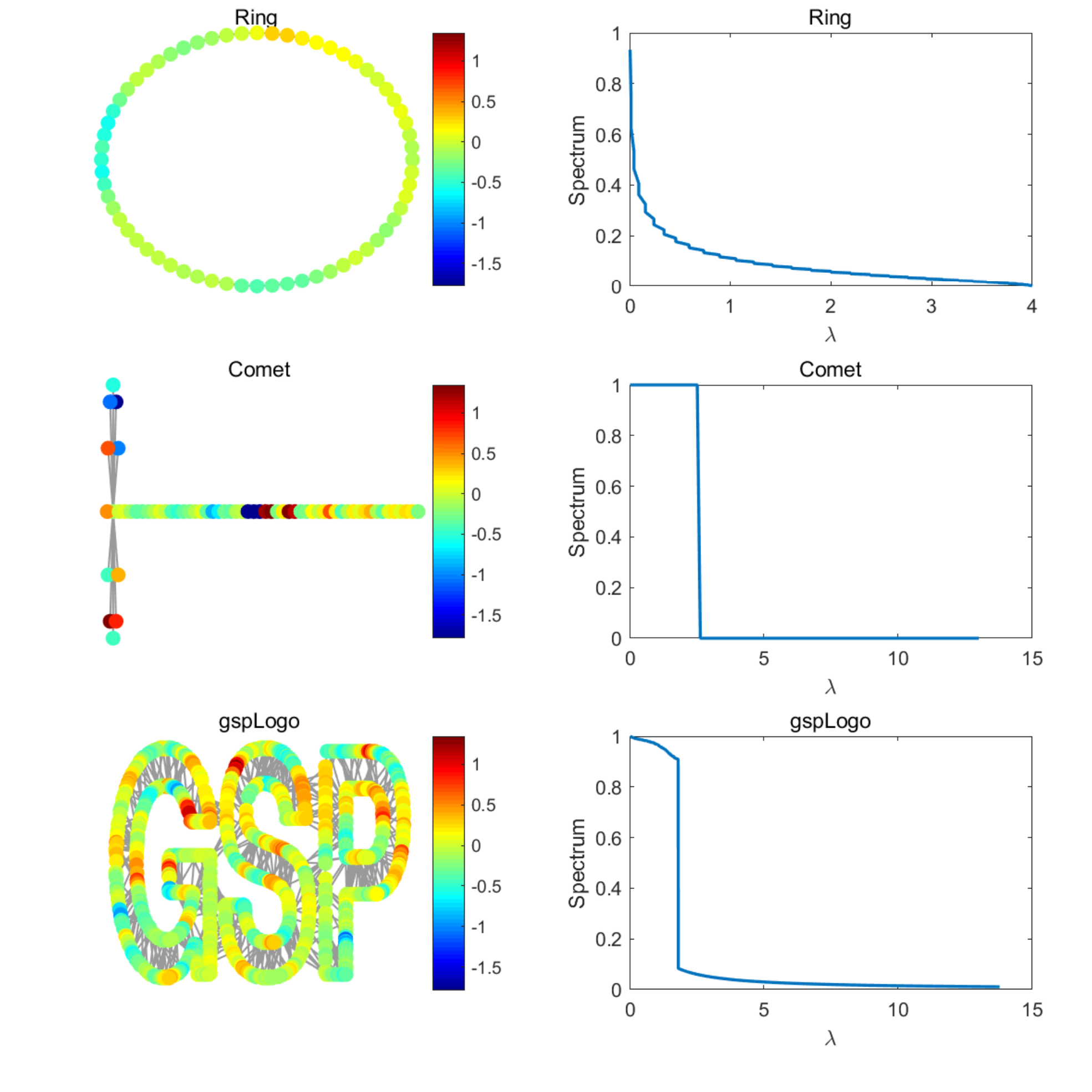}
	}
	\caption{\small The synthetic graph signals in the vertex domain and the spectral domain.}\label{fig:signals}
\end{figure}

Next, let us compare the denoising ability of the proposed method regOpt with the related methods. Experiments are performed on the ring graph with $64$ vertices, the Comet graph with $64$ vertices and the gspLogo graph with $1130$ vertices. The synthetic graph signals $\bfx$ are presented respectively in the vertex domain and the spectral domain in Figure \ref{fig:signals}. We contaminate the signals with Gaussian noise of zero mean and different standard deviations $\sigma=1/16,1/8,1/4$. In the experiment, all the LP outputs are retained and HP outputs are hard-thresholded with the value $T=3\sigma$ for reconstruction.

\begin{figure}
	\centering{
		\includegraphics[width=0.48\textwidth]{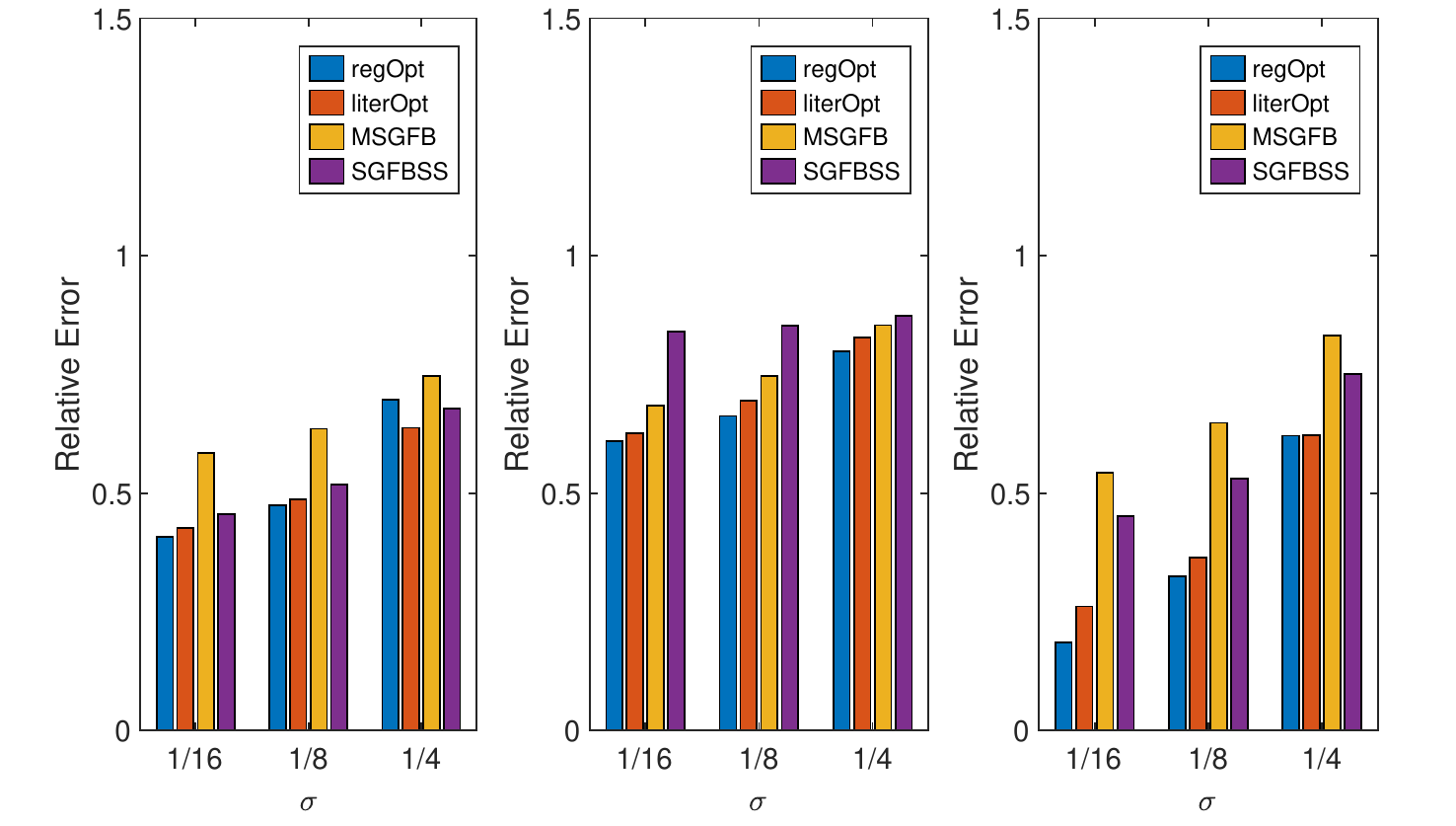}	
	}
	\caption{\small The average relative errors of $10$ runs using different models. Left: Ring graph; Middle: Comet graph; Right: gspLogo graph.}\label{fig:RE}
\end{figure}

We perform a $2$-layer decomposition for the ring graph and the Comet graph, and a $1$-layer decomposition for the gspLogo graph because SGFBSS requires $N$ to be even and there are $565$ vertices in the $2$nd layer. The hyperparameters are specified as $(r,s,J,\alpha)=(2,3,6,0.01)$ for regOpt and $J=6$ for the other models. The relative error $\frac{\|\bfy-\bfx\|_2}{\|\bfx\|_2}$ between the original signal $\bfx$ and the reconstructed signal $\bfy$ is computed, as shown in Figure \ref{fig:RE}. The results show that the proposed model regOpt outperforms the other models in most cases. 

~\\[3mm]
\section{Conclusion and Future Work}\label{sec:conclusion}
This paper describes a class of critically sampled and perfectly reconstructed spline-like filterbanks for graph signals. The analysis filters are polynomials in the normalized adjacency matrix, which allows the performance of local analysis
in the vertex domain. Besides, the lowpass filters can remove the $s$ highest frequency components of the signals, and the highpass filters can remvoe the $r$ lowest frequency components of the signals, where $r$ and $s$ are hyperparameters specified by the users.
When $r,s\ge1$, the proposed filterbanks will outperform the filterbanks proposed in the related work on denoising tasks.
	
The main limitation of the proposed filterbank is that the synthesis filter is usually not well localized. It can be challenging but rewarding to design localized synthesis filters in future work.
We also mentioned that the approximation error of the filterbank is bounded by the multiple of the largest singular value of $(\bfI_N+\bfK\bfG)^{-1}$. Empirically, we adopt a sampling pattern that prevents the smallest singular value of $\bfI_N+\bfK\bfG$ from being too small, but it may occasionally fail. In fact, this upper bound is too loose to effectively reflect the approximation error of the filterbank. As presented in the experiments on random graphs and random signals, the largest singular value of $(\bfI_N+\bfK\bfG)^{-1}$ is always much greater than the approximation errors. Thus, the future research should consider finding a tighter upper bound.

\bibliographystyle{plain}

\bibliography{ref}

\end{document}